\documentclass[twocolumn]{aastex631}

\newcommand{\jwst}{\textit{JWST}}
\newcommand{\ci}{\textsc{[C\,i]}}
\newcommand{\cii}{\textsc{[C\,ii]}}
\newcommand{\mgii}{\textsc{M}g\textsc{\,ii}}
\newcommand{\civ}{\textsc{C\,iv}}
\newcommand{\HI}{\textsc{H\,i}}
\newcommand{\Ha}{\textsc{H$\alpha$}}
\newcommand{\acii}{$\alpha_{\textsc{[Cii]}}$}
\newcommand\bbarolo{$^{\rm 3D}$\texttt{Barolo}}
\newcommand\dysmalpy{\texttt{DysmalPy}}
\newcommand\casa{\texttt{CASA}}
\newcommand{\targa}{J2318$-$3029}
\newcommand{\targb}{P009$-$10}

\newcommand{\fdm}{$f_{\rm DM}(\rm R<R_e)$}
\newcommand{\jybkms}{$\rm Jy\,beam^{-1}\,km\,s^{-1}$}
\newcommand{\kms}{$\rm km\,s^{-1}$}

\newcommand{\um}{$\mu\rm m$}
\newcommand{\Fujimoto}{Fujimoto et al. (in prep.)}

\shorttitle{Rotation curves of two $z\sim 6$ quasars}
\shortauthors{Fei et al.}
\graphicspath{{./}{figures/}}

\begin{document}


\title{Assessing the dark matter content of two quasar host galaxies at $z\sim6$ through gas kinematics}

\author[0000-0001-7232-5355]{Qinyue Fei}
\email{qyfei@pku.edu.cn}
\affiliation{Kavli Institute for the Physics and Mathematics of the Universe (Kavli IPMU, WPI), UTIAS, Tokyo Institutes for Advanced Study, University of Tokyo, Chiba, 277-8583, Japan}
\affiliation{Department of Astronomy, School of Physics, Peking University, Beijing 100871, P. R. China}
\affiliation{Kavli Institute for Astronomy and Astrophysics, Peking University, Beijing 100871, P. R. China}
\author[0000-0002-0000-6977]{John D. Silverman}
\affiliation{Kavli Institute for the Physics and Mathematics of the Universe (Kavli IPMU, WPI), UTIAS, Tokyo Institutes for Advanced Study, University of Tokyo, Chiba, 277-8583, Japan}
\affiliation{Department of Astronomy, School of Science, The University of Tokyo, 7-3-1 Hongo, Bunkyo, Tokyo 113-0033, Japan}
\affiliation{Center for Data-Driven Discovery, Kavli IPMU (WPI), UTIAS, The University of Tokyo, Kashiwa, Chiba 277-8583, Japan}
\affiliation{Center for Astrophysical Sciences, Department of Physics \& Astronomy, Johns Hopkins University, Baltimore, MD 21218, USA}
\author[0000-0001-7201-5066]{Seiji Fujimoto}\altaffiliation{Hubble Fellow}
\affiliation{Department of Astronomy, The University of Texas at Austin, Austin, TX, USA}
\author[0000-0003-4956-5742]{Ran Wang}
\affiliation{Department of Astronomy, School of Physics, Peking University, Beijing 100871, P. R. China}
\affiliation{Kavli Institute for Astronomy and Astrophysics, Peking University, Beijing 100871, P. R. China}
\author[0000-0003-4956-5742]{Luis C. Ho}
\affiliation{Kavli Institute for Astronomy and Astrophysics, Peking University, Beijing 100871, P. R. China}
\affiliation{Department of Astronomy, School of Physics, Peking University, Beijing 100871, P. R. China}
\author[0000-0002-4314-021X]{Manuela Bischetti}
\affiliation{Dipartimento di Fisica, Universit\'a di Trieste, Sezione di Astronomia, Via G.B. Tiepolo 11, I-34131 Trieste, Italy}
\affiliation{INAF - Osservatorio Astronomico di Trieste, Via G. B. Tiepolo 11, I-34131 Trieste, Italy}
\author[0000-0002-6719-380X]{Stefano Carniani}
\affiliation{Scuola Normale Superiore, Piazza dei Cavalieri 7, I-56126 Pisa, Italy}
\author[0000-0002-9122-1700]{Michele Ginolfi}
\affiliation{Dipartimento di Fisica e Astronomia, Università di Firenze, Via G. Sansone 1, I-50019, Sesto F.no (Firenze), Italy}
\affiliation{INAF — Osservatorio Astrofisico di Arcetri, Largo E. Fermi 5, I-50125, Florence, Italy}
\author[0000-0002-0267-9024]{Gareth C. Jones}
\affiliation{Department of Physics, University of Oxford, Denys Wilkinson Building, Keble Road, Oxford OX1 3RH, UK}
\affiliation{Cavendish Laboratory, University of Cambridge, 19 J. J. Thomson Ave., Cambridge CB3 0HE, UK}
\affiliation{Kavli Institute for Cosmology, University of Cambridge, Madingley Road, Cambridge CB3 0HA, UK}
\author[0000-0002-4985-3819]{Roberto Maiolino}
\affiliation{Cavendish Laboratory, University of Cambridge, 19 J. J. Thomson Ave., Cambridge CB3 0HE, UK}
\affiliation{Kavli Institute for Cosmology, University of Cambridge, Madingley Road, Cambridge CB3 0HA, UK}
\affiliation{Department of Physics \& Astronomy, University College London, Gower Street, London WC1E 6BT, UK}
\author[0000-0002-0303-499X]{Wiphu Rujopakarn}
\affiliation{National Astronomical Research Institute of Thailand, Don Kaeo, Mae Rim, Chiang Mai 50180, Thailand}
\affiliation{Department of Physics, Faculty of Science, Chulalongkorn University, 254 Phayathai Road, Pathumwan, Bangkok 10330, Thailand}
\author[0000-0003-4264-3381]{N. M. {F\"{o}rster Schreiber}}
\affiliation{Max-Planck-Institut f\"{u}r extraterrestrische Physik (MPE), Giessenbachstr. 1, D-85748 Garching, Germany}
\author[0000-0001-6703-4676]{Juan M. Espejo Salcedo}
\affiliation{Max-Planck-Institut f\"{u}r extraterrestrische Physik (MPE), Giessenbachstr. 1, D-85748 Garching, Germany}
\author[0000-0001-7457-4371]{Lilian L. Lee}
\affiliation{Max-Planck-Institut f\"{u}r extraterrestrische Physik (MPE), Giessenbachstr. 1, D-85748 Garching, Germany}

\begin{abstract}
We conduct a study of the gas kinematics of two quasar host galaxies at $z\gtrsim6$ traced by the \cii\ emission line using ALMA. By combining deep observations at both low and high resolution, we recover the diffuse emission, resolve its structure, and measure the rotation curves from the inner region of the galaxy to its outskirts using \dysmalpy\ and \bbarolo. Assuming that both galaxies exhibit disk rotation driven by the gravitational potential of the galaxy, we find that the best-fit disk models have a $V_{\rm rot}/\sigma \approx 2$ and inferred circular velocities out to $\sim$6-8 kpc scales, well beyond the likely stellar distribution. 
We then determine the mass profiles of each component (stars, gas, dark matter) with priors on the baryon and dark matter properties. We find relatively large dark matter fractions within their effective radii (\fdm = $0.61_{-0.08}^{+0.08}$ and $0.53_{-0.23}^{+0.21}$, respectively), which are significantly larger than those extrapolated from lower redshift studies and remain robust under different input parameters verified by Monte-Carlo simulations. The large \fdm ~corresponds to halo masses of $\sim 10^{12.5}-10^{12.8}\, M_\odot$, thus representative of the most massive halos at these redshifts. Notably, while the masses of these SMBHs are approximately 1 dex higher than the low-redshift relationship with stellar mass, the closer alignment of SMBH and halo masses with a local relationship may indicate that the early formation of these SMBHs is linked to their dark matter halos, providing insights into the co-evolution of galaxies and black holes in the early universe. 

\end{abstract}



\section{Introduction} 
\label{sec1: intro}

Galaxy formation and evolution are initiated by dark matter assembly within the $\Lambda$CDM framework \citep{Colberg+2000, Springel+2005}. In this framework, the most massive dark matter halos in the earliest epochs of the universe correspond to the largest fluctuations in cosmic density fields, providing key constraints on the $\Lambda$CDM model. Therefore, it is crucial to study the dark matter properties in the first billion years of the universe ($z\gtrsim 6$).

Observations of dark matter (DM) assembly have been achieved through detailed kinematics studies for a few decades \citep[e.g., ][]{Rubin+1978, Rubin+1980, Corbelli+2000, Sofue+2001, deBlok+2008, Genzel+2017, Genzel+2020, Rizzo+2021}. This involves the examination of rotation curves (RCs; rotation velocity versus radius) of nearby galaxies using cold gas tracers such as \HI\ and CO \citep[e.g., ][]{Walter+2008, Levy+2018, Lang+2020}, as well as higher redshift galaxies using \Ha\ and \cii\ \citep[e.g., ][]{Wuyts+2016, Genzel+2006, Genzel+2011, Genzel+2017, Genzel+2020, Tiley+2019, Rizzo+2021, Rizzo+2023, Parlanti+2023, Puglisi+2023}. Previous studies of low redshift galaxies using \HI\ kinematics have shown, at the outskirts of the galaxies ($r\gg R_e$ where $R_e$ is the effective radius of the stellar disk), that rotation velocities remain constant with radius thus resulting in a flattened RC which indicates the existence of dark matter \citep[e.g.,][]{Persic+1996}. 

However, studies focusing on massive disk galaxies at higher redshift present a more complicated scenario. It has been recently observed that the RCs of galaxies at $z\sim 2$ decrease significantly at their outskirts \citep{Lang+2017}, suggesting that massive, star-forming galaxies at high redshift are baryon-dominated systems on galactic-scale \citep{Wuyts+2016, Genzel+2017, Genzel+2020}. In addition, the fraction of dark matter within the effective radius (\fdm) decreases with increasing redshift \citep{Genzel+2020, NestorShachar+2023}. In contrast, declining RCs have not been found in several studies \citep[e.g., ][]{Tiley+2019} that probed a different galaxy parameter space, used a different methodology, and, importantly, considered DM fractions within an appreciably larger region $\lesssim~3.5 R_e$.

Investigating the distribution of dark matter across cosmic time is fundamental to understand galaxy formation and evolution. Indeed, galaxies were initially formed within dark matter halos and then grew by accreting gas from their environments \citep{Keres+2005, Dekel+2006, Dekel+2009} and/or through mergers \citep{Kauffmann+1993, Mihos+1996, Kormendy+2009}. The rapid radial movement of baryons, including gas inflows and/or outflows, can shape the formation of galaxies and also impact the properties of their dark matter halos at the same time \citep{Dekel+1986, El-Zant+2001, Goerdt+2010, Cole+2011, Martizzi+2013, Ferundlich+2020, Dekel+2022}. The combination of dynamical friction from incoming baryonic clumps and feedback from supernovae and AGNs causes baryons and dark matter to be expelled which may lead to a dark matter deficit in the core of massive galaxies at $z\gtrsim 2$ \citep{Genzel+2017, Genzel+2020, ForsterSchreiber+2018, ForsterSchreiber+2019, Price+2021, NestorShachar+2023}. 

Similar studies have been conducted at higher redshifts (e.g., \citealt{Rizzo+2020, Rizzo+2021, Fraternali+2021, Lelli+2021, Neeleman+2021, Neeleman+2023, RomanOliveira+2023, Fujimoto+2024, Rowland+2024}). These high-resolution observations mainly focus on the disk-scale gas kinematics that provide an accurate description of their RCs in the inner regions but not probing the outskirts where the flattened and/or falling-off part of the RCs is most sensitive to dark matter. In addition, those studies focus on galaxies with relatively lower stellar mass or massive, compact starbursts well above the main sequence, and thus cannot be directly compared with massive galaxies at cosmic noon. In order to better understand the $\Lambda$CDM model and galaxy evolution models, it is important to investigate the dark matter assembly of massive galaxies ($\log M_\star/M_\odot\gtrsim 10$) at higher redshift. 

To find such massive galaxies, we can use the luminous quasars, powered by accretion onto supermassive black holes (SMBHs), as signposts since their host galaxies are likely to have high stellar masses given their massive black holes ranging from $10^8-10^{10}\,M_\odot$ and have been observed up to $z\gtrsim 7$ \citep{Banados+2018, Yang+2020, Wang+2021, Ding+2023}. In fact, recent studies confirmed the remarkable galaxy overdensity around luminous quasars at $z\gtrsim 6$, favoring the scenario that such quasars reside in massive dark matter halos \citep[e.g.,][]{Kashino+2023, Wang+2023, Eilers+2024, RojasRuiz+2024}. 

ALMA studies have identified extended \cii\ 158 \um\ line structures with $r\gtrsim10\,\rm kpc$ around several high-$z$ quasars at $z\sim 6-7$ \citep{Cicone+2015, Bischetti+2024}, which are well beyond the galactic scales even in the most massive galaxies at these redshifts \citep{Yang+2022, Morishita+2024}. The kinematics of this extended gas may be used to examine the dark matter contribution similar to the local \HI\ studies (e.g., \citealt{Corbelli+2000, Walter+2008}). Additionally, the galaxy masses of these quasars are expected to be massive according to the local $M_{\rm BH}-M_\star$ relation \citep{Kormendy+2013, Ding+2022a, Tanaka+2024}, which makes them an ideal sample for comparison with massive galaxies at $z\sim 2$ \citep{Genzel+2020, NestorShachar+2023}. Therefore, it is important to study the gas kinematics of quasars at $z\sim 6$, particularly focusing on the outskirts of their host galaxies to assess the contribution of dark matter. 

Here, we use \cii\ 158 \um\ to trace the kinematics of the gas out to $\sim8$ kpc for the host galaxies of two quasars (\targa\ and \targb) identified out of a sample of five at $z>6$. Their RCs enable us to model the contributing components and determine the dark matter content for the first time at these high redshifts. This enables us to measure the dark matter fraction, halo mass, and relation to the black hole mass. The paper is organized as follows. Section \ref{sec2: data} summarizes the data selection and the data reduction process. Section \ref{sec3: methods} introduces the two kinematic analysis methods used in this work. Section \ref{sec4: results} presents the results of the analysis, including the kinematic fitting results and the parameters of interest. The direct measurement and comparison with previous studies are also shown. Section \ref{sec5: discussion} discusses the robustness of our measurements of the dark matter fraction and the dark matter halo mass, and presents a comparison with that measured from different methods. The summary and conclusion are shown in Section \ref{sec6: conclusion}. The cosmological parameters of $\Omega_{\rm m}=0.308$, $\Omega_\Lambda=0.692$, and $H_0=67.8\,\rm km\,s^{-1}\,Mpc^{-1}$ are used in the analysis \citep{Planck+2016}, resulting in 1 arcsec equal to 5.84\,kpc at $z=6$.


\section{Data}
\label{sec2: data}
\begin{deluxetable*}{ccccccccc}
    \tablenum{1}
    \tablecaption{ALMA observations \label{Tab1: observation}}
    \tablewidth{0pt}
    \tablehead{
    \colhead{Name} & \colhead{$z_\cii$} & \colhead{Cycle} & \colhead{Configuration} & \colhead{On Source} & \colhead{Synthesized} & \colhead{Spectral} & \colhead{Project} & \colhead{Reference} \\
    \colhead{} & \colhead{} & \colhead{} & \colhead{} & \colhead{Time} & \colhead{Beam} & \colhead{Resolution} & \colhead{ID} & \colhead{} \\
    \colhead{} & \colhead{} & \colhead{} & \colhead{} & \colhead{(min)} & \colhead{} & \colhead{($\rm km\,s^{-1}$)} & \colhead{} & \colhead{}
    }
    \decimalcolnumbers
    \startdata
    \targb & 6.0040 & 5 & C43-5 & 26.6 & $0\farcs32\times 0\farcs22$ & 4.28 & 2017.1.01301.S & \cite{Venemans+2020} \\
          & & 8 & C4 & 60.8 & $0\farcs46\times0\farcs34$ & 17.26 & 2021.1.01320.S & \Fujimoto \\
          & & 8 & C1 & 20.7 & $1\farcs27\times0\farcs91$ & 17.26 & 2021.1.01320.S & \Fujimoto \\
    \targa & 6.1456 & 3 & C36-2 & 7.08 & $0\farcs86\times0\farcs74$ & 4.39 & 2015.1.01115.S & \cite{Decarli+2018} \\
          & & 6 & C43-6 & 40.5 & $0\farcs10\times0\farcs10$ & 4.42 & 2018.1.00908.S & \cite{Venemans+2020} \\
          & & 8 & C4 & 113.9 & $0\farcs45\times0\farcs34$ & 17.68 & 2021.1.01320.S  & \Fujimoto \\
          & & 8 & C1 & 39.0 & $1\farcs18\times0\farcs93$ & 17.68 & 2021.1.01320.S & \Fujimoto \\
    \enddata
    \tablecomments{(1) Name of targets; (2) \cii\ redshift adopted from \citet{Venemans+2020}; (3) ALMA cycle; (4) Corresponding configuration; (5) Total on-source time; (6) The FWHM of the synthesized beam, with robust=0.5; (7) The original spectral resolution; (8) The project ID of corresponding observation; (9) Reference.}
\end{deluxetable*}

\begin{deluxetable*}{ccccc}
    \tablenum{2}
    \tablecaption{\cii\ data cubes \label{Tab2: cubes}}
    \tablewidth{0pt}
    \tablehead{
    \colhead{Name} & \colhead{Syn. Beam} & \colhead{P.A.} & \colhead{rms noise} & \colhead{Chan. Width} \\ 
    \colhead{} & \colhead{} & \colhead{$(^\circ)$} & \colhead{$(\rm mJy\,beam^{-1})$} & \colhead{$(\rm km\,s^{-1})$}
    }
    \decimalcolnumbers
    \startdata
    \targb(high-res) & $0\farcs35\times 0\farcs29$ & $-89.39$ & 0.22(0.37) & 34.52 \\
    \targb(low-res) & $0\farcs72\times 0\farcs63$ & $74.71$  & 0.12(0.19) & 34.52 \\
    \targa(high-res) & $0\farcs22\times 0\farcs16$ & $-72.01$ & 0.097(0.26) & 35.35 \\
    \targa(low-res) & $0\farcs71\times 0\farcs66$ & $-81.29$ & 0.098(0.13 )& 35.35 \\
    \enddata
    \tablecomments{(1) Name of targets; (2) Synthesized beam size of the final data cube; (3) The position-angle of the corresponding data cubes; (4) The root mean square (rms) noise of the data cube after applying the JvM correction, with the pre-correction rms level shown in parentheses; (5) The width of each channel of data cubes. The high-resolution data for \targb\ is a combination of Cycle5 and Cycle8 C4 observation. The low-resolution data for this target is a combination of Cycle 3 and Cycle 8 observations. The high-resolution data for \targa\ is combined from Cycle 6 observation and Cycle 8 C4 data.}
\end{deluxetable*}

\begin{figure*}
    \centering
    \includegraphics[width=0.49\linewidth]{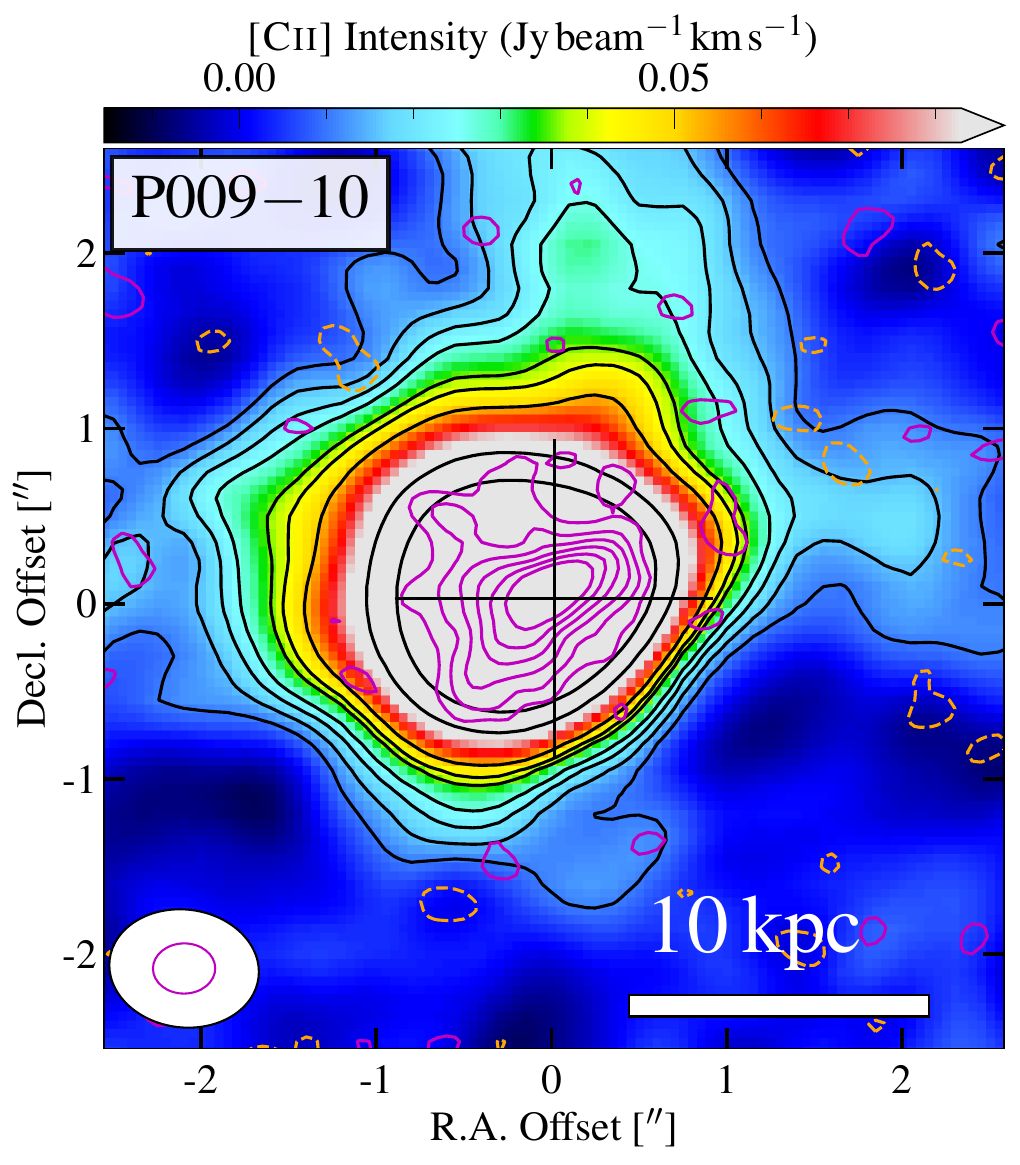}
    \includegraphics[width=0.49\linewidth]{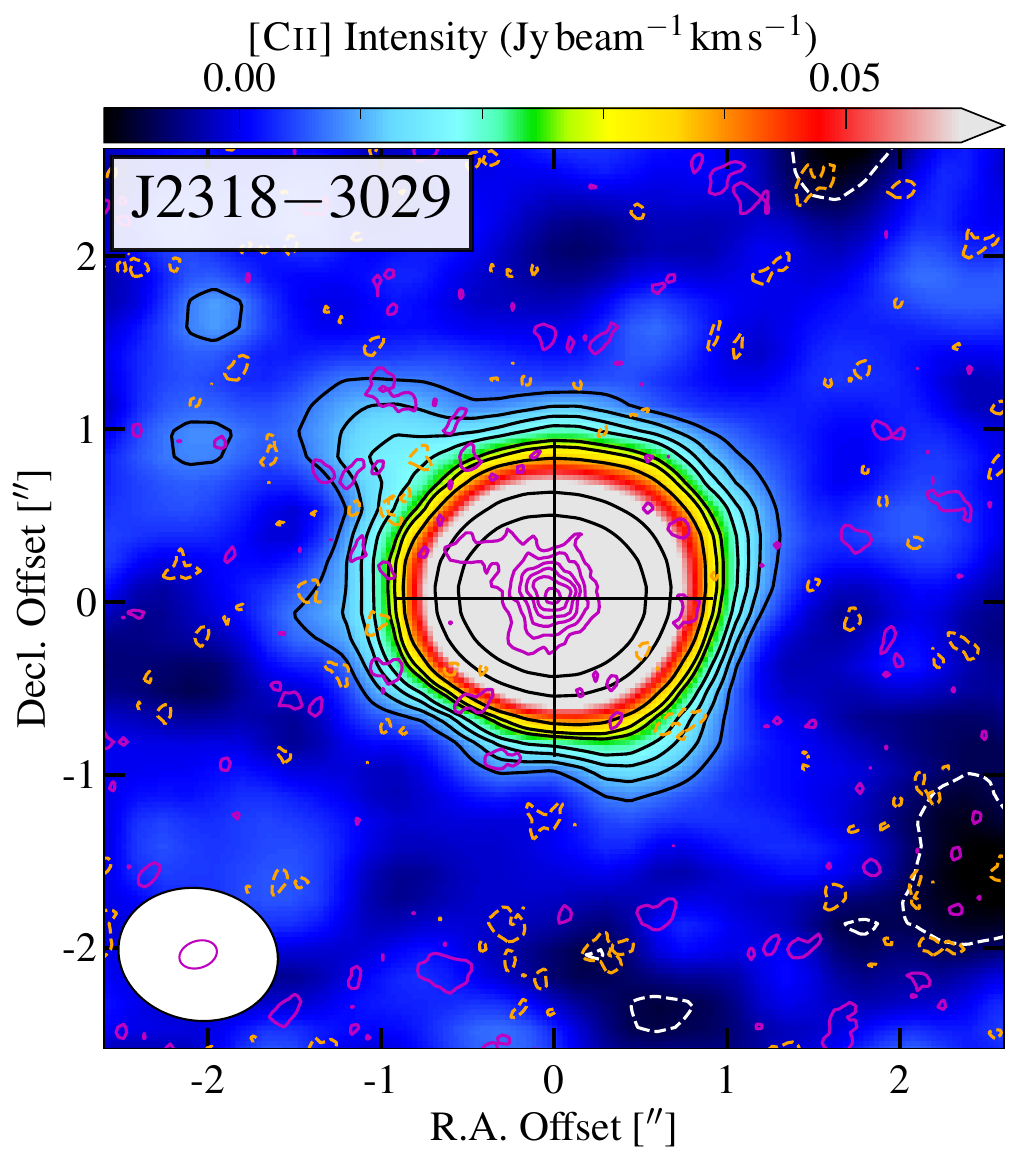}
    \caption{\cii\ intensity maps of \targb\ (left) and \targa\ (right) for the low-resolution data. Black contours represent the 2,3,4,5,7,8,10,20,30 $\times$ $\sigma$, where $\sigma$ is the root mean square (rms) noise of the line-free region in the corresponding intensity map. Magenta contours represent the \cii\ intensity of the high-resolution data, at levels of 2,4,6,8,10,12$\times$$\sigma$. The white and orange dashed curves represent the -2$\sigma$ of low and high-resolution data. The synthesized beams for both low- and high-resolution data are depicted as white and magenta ellipses at the bottom left corner of each panel. A 10\,kpc scale bar is shown in the left panel and applies to the right panel as well. The quasar location, as adopted from \citet{Venemans+2020}, is marked by a black cross.}
    \label{Fig1: intensity map}
\end{figure*}

We selected five quasars that showed initial signs of extended \cii\ structures in both visibility- and image-based analyses (see Fujimoto et al. in prep. for further details), from a systematic analysis of the public ALMA archive which includes $\sim$40 quasars at $z \sim 6-7$ \citep{Decarli+2018, Novak+2020}. All five quasars show a possible enhancement of the line flux at the shortest baselines, which corresponds to an angular scale of $\theta\sim 10''$ ($\sim$60\,kpc in radius at $z=6$) and significantly higher than the best-fit single Gaussian by $\sim$2–3 $\rm Jy\,km\,s^{-1}$. The $uv-$tapered maps confirm that the \cii\ radial profiles extend beyond the ALMA beams and the dust continuum maps. These five quasars offer the best opportunity to explore the CGM-scale study around massive galaxies in the early Universe in detail with ALMA. We subsequently performed deep ALMA follow-up using both compact (C1) and moderately extended (C4) configurations to recover the extended \cii\ emission as well as resolve its structure for these five quasars (\#2021.1.01320.S; PI: J. Silverman; Fujimoto et al. in prep.).

In our analysis, we focus on scales within a radius of $\sim 10\,\rm kpc$, where the signal-to-noise ratio (SNR) of \cii\ emission is high and enables secure measurements of the gas kinematics for two quasars (\targb\ and \targa) among the five (Sec. \ref{sec3: methods}). We remove J0100$+$2802 and P308$-$21 since their velocity fields are highly perturbated by close companions thus likely to be merging systems \citep{Tripodi+2024, Decarli+2019, Loiacono+2024}. We also remove P359$-$06 from our study since the kinematic analysis suggests a dispersion-dominated system, which cannot be described by a rotating disk model. We include all available data from the ALMA archive related to our designated targets. In order to optimize the kinematic information provided by the different ALMA configurations, we create data cubes of the \cii\ emission at both high and low resolutions using the common astronomy software application \citep[\casa; ][]{McMullin+2007} through the amalgamation of various ALMA configurations. The details of the observations used for generating the \cii\ data cubes are listed in Table \ref{Tab1: observation}. 

We calibrated the data employing the corresponding version of \casa\ and then used version 6.4.2 for the cleaning process and subsequent creation of final data cubes. Using the \casa\ task \texttt{concat}, we combined previous high-resolution observations with our C4 configuration data and merged existing low-resolution observations with our new observations to obtain the emission on galactic scales and further out. We subtracted the continuum for both visibilities using the \casa\ task \texttt{uvcontsub} with a first-order polynomial fit to the line-free channel. The final \cii\ data cubes were generated from the continuum-subtracted visibility data.

We used the \texttt{tclean} task in \casa\ to produce high-resolution images. The \cii-line data cube was cleaned using Briggs weighting (for high-resolution data, we used uniform weighting) and a stop threshold set at 2.0 times the root-mean-square (rms) noise of the off-source channels. During \texttt{tclean}, we specified the gridder as mosaic and utilized the auto-multithresh masking procedure as outlined in \citep{McMullin+2007}. The parameters of masking were set based on CASA guidelines.\footnote{\url{https://casaguides.nrao.edu/index.php?title=Automasking\_Guide\_CASA\_6.5.4}}

For the low-resolution data, we cleaned the data cube using the same channels of high-resolution data, with the $\texttt{robust}=2.0$ in order to capture the extended emission as deep as possible, and an additional $\texttt{uvtaper}$ of $0\farcs3$ was applied in the image domain to increase the SNR of the extended \cii\ emission. The other clean parameters are the same as those used in generating high-resolution data. Realizing that the shape of the dirty beam is not an ideal Gaussian-like pattern since the combination of different observations results in non-uniform coverage of the $uv$ plane, we applied the Jorsater \& van Moorsel correction \citep[][JvM correction]{Jorsater+95, Czekala+21} to the cleaned data and performed all analyses on these final data products. In Appendix \ref{secA}, we describe the correction procedure in further detail. By comparing the data products before and after applying the JvM correction, we find that approximately 9\% of \cii\ flux for \targb\ and 4\% for \targa\ are reduced for the low-resolution data. 
The basic information about the data cubes after applying the JvM correction is listed in Table \ref{Tab2: cubes}, and the comparison between the data products before and after applying the JvM correction is shown in Appendix \ref{secA} and Fig. \ref{fig13: JvM}. 

Figure \ref{Fig1: intensity map} shows the intensity maps of the \cii\ emission for \targb ~and \targa ~generated from the low-resolution data and the \casa\ task \texttt{immmoments}. We collapsed all channels within 1.7 times the full width at half maximum (FWHM) centered on the peak of the \cii\ line for each source. Our purpose is to show the extended \cii\ emission while maximizing the SNR. We also created \cii\ intensity maps of the high-resolution data using the same frequency range, which are also depicted in Figure \ref{Fig1: intensity map}. The rms noise levels for the low-resolution \cii\ maps of \targb\ and \targa\ are 0.045 and 0.022\,\jybkms, respectively, while the rms noise levels are 0.073 and 0.040\,\jybkms for the high-resolution images. For both, it is evident from the low-resolution data that the structure is well resolved by the beam size which is well beyond the typical stellar size of the galaxy or quasar hosts at similar redshifts \citep[$\sim 1-2\,$kpc][]{Shibuya+2015, Morishita+2024}.

\section{Analysis methods}
\label{sec3: methods}
Numerous methods and software tools have been developed to investigate gas kinematics including non-parametric and parametric models for galaxies across a wide range of redshift \citep[e.g., ][]{Rogstad+1974, Jozsa+2007, DiTeodoro&Fraternali2015, Bouche+2015, Genzel+2017, Rizzo+2021}. In this context, `non-parametric' refers to a fitting approach that characterizes only the kinematic properties, such as rotational velocity and velocity dispersion, without assuming an underlying mass distribution model. In contrast, a parametric model aims to explore the mass distribution by solving Newton's law of gravity to fit the observed gas kinematics.

In this study, we primarily use \dysmalpy\ to examine the gas kinematics in two quasar hosts and analyze their mass distribution \citep[e.g., ][]{Price+2021}. Additionally, we use \bbarolo \citep{DiTeodoro&Fraternali2015}, which is a non-parametric kinematic model, to directly estimate the rotation velocities independently of the model. Assuming that the \cii\ emission from our targets primarily originates from a rotating gaseous disk, with the gas motion governed by the gravitational potential, we employed the aforementioned kinematic analysis tools to model the \cii\ kinematics and derive the mass distribution. However, as a word of caution, we identified significant non-circular motions in \targb, which impact the accuracy of the kinematic analysis. As a result, we remove this non-circular component before applying the standard kinematic analysis to the optimized data as described below.

\subsection{Non-circular motion filtering in \targb}
\label{sec3.3: removing non-circular gas}
\begin{figure*}
    \centering
    \includegraphics[width=\linewidth]{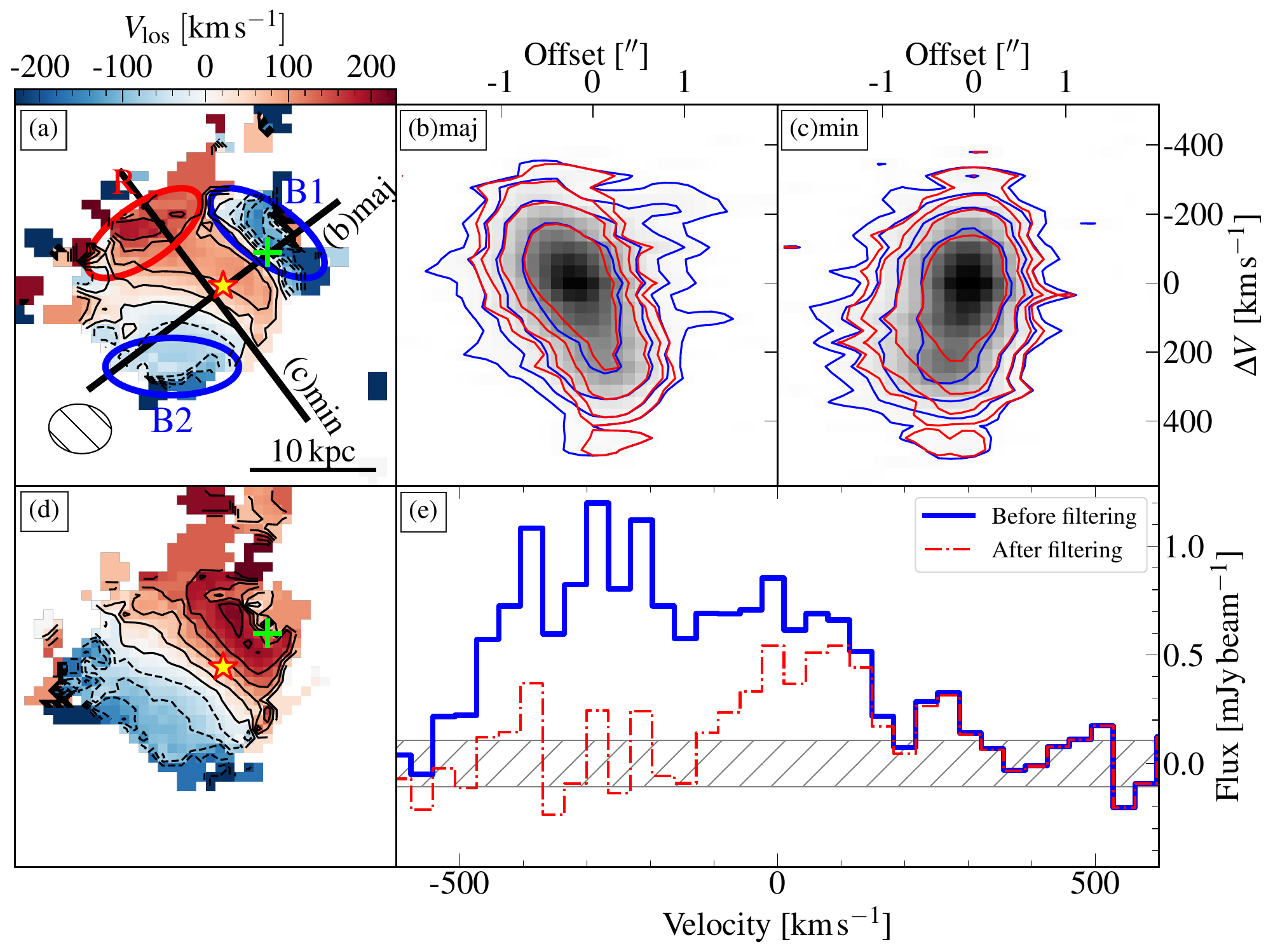}
    \caption{Comparing the \cii\ emission before and after filtering. {\it Panel (a) and (d):} The line-of-sight velocity map generated from the original data cube (a) and after applying the filter (d). {\it Panel (b) and (c):} Position-Velocity (PV) diagram extracted from the major- and minor-axis of the original cube (blue contours) and after filtering application (red contours). The major- and minor-axis are shown as the black line in panel (a). {\it Panel (e):} An example of spectra before and after filtering out the external emission, extracted from the pixel-wise region indicated by the green cross. The red star in panels (a) and (d) indicates the location of the quasar \citep{Venemans+2019}.}
    \label{fig2: filtering}
\end{figure*}

While both models are based on the assumption that the gas motion adheres to regular rotational dynamics, we note that the original line-of-sight (los) velocity map (Figure~\ref{fig2: filtering}a) reveals an anomalous \cii\ kinematic feature, characterized by two blue-shifted components (regions B1 and B2 in Figure \ref{fig2: filtering}) and one red-shifted component (region R in Figure \ref{fig2: filtering}). Such kinematic anomalies in quasar host galaxies may arise from strong quasar feedback \citep{Feruglio+2010, Maiolino+2012, Harrison+2014}, newly accreted gas from satellite galaxies \citep{Dekel+2009, Bischetti+2024}, or a combination of both mechanisms \citep{Fujimoto+2019, Triopodi+2024}. To isolate the well-settled disk, observed through kpc-scale observations \citep{Neeleman+2021}, that is solely influenced by gravitational potential, we filtered out any non-circular motions.

To accomplish this, we constructed a rotating disk model using high-resolution data and convolved it to match the angular resolution of the low-resolution data. The high-resolution data sets are presumably tracing the galactic-scale kinematics and are not sensitive to the non-circular motions that more commonly appear on large scales. Therefore by comparing the smoothed data with intrinsic low-resolution data sets, it is easier to identify such non-circular motions. By subtracting the convolved model from the low-resolution observations, we generated a residual spectrum for each pixel. We then fit each residual spectrum with two models, a Gaussian profile and a constant, using \texttt{lmfit}.\footnote{\url{https://lmfit.github.io/lmfit-py}} We compared the Bayesian Information Criterion (BIC) for both fitting results. If the BIC for the Gaussian fit was smaller than that for the constant, we considered the non-circular motion at that pixel to represent a `real' emission rather than an artifact. For these pixels, we subtracted the Gaussian best-fit result from the data cube.

The results of this filtering process are illustrated in Figure~\ref{fig2: filtering}d. The remaining \cii\ velocity structure, after removing the non-circular components, exhibits a regular rotating disk with a clear velocity gradient. The effectiveness of this filtering is further demonstrated in the position-velocity (PV) diagram (panels b and c). A comparison of the PV-map from both the original and optimized data shows that the external components have been removed, while the primary rotating disk remains intact. Figure~\ref{fig2: filtering}e presents an example of spectra extracted from both the original and optimized data.

The optimized data cube is considered to represent the \cii\ disk, primarily governed by gravitational potential, with minimal non-circular motion. It is noteworthy that this method is particularly significant for \targb, while it has minimal impact on \targa\ due to the absence of similarly non-circular components. Consequently, the original data cube of \targa\ was used in the kinematic analysis.

\subsection{\dysmalpy}
\label{sec3.1: dysmalpy}
The Python-based \dysmalpy,\footnote{\url{https://www.mpe.mpg.de/resources/IR/DYSMALPY}}, or its parent IDL version \texttt{Dysmal}, is a versatile forward-modeling tool. It is based on multi-component mass models and has a long history of development. This tool has been used in near-IR/optical IFU and millimeter interferometric studies of disk galaxies at high-$z$ \citep[e.g., ][]{Genzel+2006, Genzel+2011, Genzel+2017, Genzel+2020, Cresci+2009, Wuyts+2016, Burkert+2016, Lang+2017, Tadaki+2017, Ubler+2018, Ubler+2019, Ubler+2024, Price+2021, HerreraCamus+2022, NestorShachar+2023, Lee+2024} and low-$z$ (e.g., \citealt{Davies+2004a, Davies+2004b, Davies+2009, Davies+2011, Davies+2014, Sani+2012, MullerSanchez+2013, Lin+2016}). A detailed description of \dysmalpy's model construction and optimization is given by \citet{Price+2021}. In general, \dysmalpy\ is based on a mass distribution from which the kinematics are computed, using the following equation:
\begin{equation}
    v_c^2(R) = \sum_i \left[\frac{1}{R}\frac{\partial \Phi_i}{\partial R}\right] + \frac{1}{R}\frac{\partial p}{\partial R}.
\end{equation}
The first term in the right hand of the equation showcases the gravitational potential, which is summed up by the baryon (stars and gas) and dark matter component, and the second term shows the effect of gas pressure \citep[$p \propto \sigma^2$; asymmetric drift correction;][]{Burkert+2010}. The velocity dispersion is assumed to be locally isotropic and radially uniform, representing a dominant turbulence term $\sigma$. 

With the above-mentioned assumptions, \dysmalpy\ generates the intrinsic composite model as a 4D hypercube, summing up the components accounting for projection according to the inclination angle ($i$), position angle ($\phi$), and relative flux weighting. Each cell of the hypercube contains the total model flux in the `sky' coordinates ($x_{\rm sky}, y_{\rm sky}, z_{\rm sky}$) and its full line-of-sight velocity distribution, which is then collapsed along $z_{\rm sky}$ and convolved with a 3D kernel folding in the spatial point spread function (PSF) and spectral line spread function (LSF). This procedure accounts for beam smearing, velocity resolution, and broadening of the line-of-sight velocity distribution due to projection effects.

A typical massive galaxy can consist of four components: a central supermassive black hole (SMBH), a stellar bulge, a gaseous disk, and a dark matter halo \citep[e.g.,][]{Lelli+2021, Tripodi+2024}. 
While we acknowledge that the gravitational potential of SMBHs with masses of $\sim 10^9\,M_\odot$ may play a substantial role in our targets \citep{Neeleman+2021}, the sphere of influence (SoI) of these SMBHs is expected to be minimal when accounting for the mass contribution from baryonic matter. High-resolution observations of quasars at comparable redshifts have demonstrated an SoI radius of $\lesssim 100\,$pc \citep{Walter+2022}, considerably smaller than our spatial resolution. Therefore, we omit the SMBH contribution in our fitting process. 

We have assigned the total baryonic mass as a free parameter, with an initial estimate of $\log (M_{\rm bar}/M_\odot)=10.5$. For our modeling, we chose to assign all stellar mass to a bulge component and assume that all the gas mass is distributed in a disk component as traced by the \cii~ light. While \citet{Tacconi+2020} suggests that the ratio of stellar mass to total baryonic mass ($f_\star$) is approximately 0.2 at $z\gtrsim 6$, this value can exhibit significant variation for individual galaxies. Consequently, we treat $f_\star$ as a free parameter in our fitting process, applying a flat prior with a range between 0 and 1 to account for this potential scatter. 

We use an oblate spheroid to describe the 3D mass distribution of the stellar component, which can be characterized by a deprojected S\'ersic profile with the following parameters: the stellar mass ($M_\star$), the effective radius ($R_\star$), the S\'ersic index ($n_\star$), and the inverse axis ratio representing the ratio between the major axis and the height ($q_\star^{-1}$). The $R_\star$ is a free parameter during the fitting since we have no direct observation of the stellar light from our target. We found that $n_\star$ cannot be left entirely free during fitting, as extreme values (e.g., $n_\star<0.5$) can cause the fitting process to fail. Therefore, we impose strong priors on these parameters. Based on observations of stellar light from quasar host galaxies at similar redshifts \citep{Ding+2023}, we fix $n_\star = 1.0$ in the fitting. For simplicity, we set $q_\star^{-1} = 1$ for the stellar component, assuming it forms an isotropic sphere. The stellar mass is directly linked to the $M_{\rm bar}$ with $f_\star$. 

We use the same function to describe the gas distribution, with the parameters for the gas mass ($M_{\rm gas}$), the effective radius ($R_{\rm gas}$), the S\'ersic index ($n_{\rm gas}$), and the inverse axis ratio ($q_{\rm gas}^{-1}$). 
While \cii\ has been proven to be a good tracer of molecular gas \citep[e.g., ][]{Zanella+2018, Madden+2020, Vizgan+2022}, the conversion is dependent on galaxy properties \citep[e.g., ][]{Khatri+2024}. Therefore, we fix ngas to the value derived through a fit to the integrated \cii\ map (see Section 4.1), but leave $R_{\rm gas}$ free.
We adopt $q_{\rm gas}^{-1} = 5$, assuming a morphology more similar to disk galaxies at $z \sim 2$ \citep{vanderWel+2014}. Similarly, the gas mass is related to the total baryonic mass and the stellar mass fraction. 

Those parameters (e.g., $n_\star$, $q_\star^{-1}$, $q_{\rm gas}^{-1}$, etc) without direct constraints from the observation may introduce significant uncertainties in our fitting. Therefore, we further discuss the influences of adopting different values for these parameters in Section \ref{sec5.4: possibilities}. The other geometric parameters related to the kinematics, e.g., the inclination angle and the position angle, are free parameters in the fitting.

Accounting for the dark matter content, we fit the mass ratio between the dark matter mass enclosed within the effective radius and the dynamical mass in this region (dark matter mass fraction) [\fdm] instead of fitting the virial mass of the dark matter halo. This parameter has an initial estimate of 0.5 and a flat prior between 0 and 1. The virial mass of the dark matter halo is connected to the \fdm. The concentration is fixed at 3.5, which is expected at this redshift \citep{Dutton+2014}.

To summarize, we model the gas kinematics using \dysmalpy\ with the following parameters for mass distribution,
\begin{itemize}
    \item Baryonic mass ($M_{\rm bar}$).
    \item Stellar mass fraction ($f_\star$).
    \item Effective radius of the stellar component ($R_\star$).
    \item S\'ersic index of the stellar component ($n_\star$) that is fixed to 1.
    \item Inverse axis ratio of the stellar component ($q_\star^{-1}$) that is fixed to 1.
    \item Effective radius of the gas component ($R_{\rm gas}$).
    \item S\'ersic index of the gas component ($n_{\rm gas}$).
    \item Inverse axis ratio ($q_{\rm gas}^{-1}$) that is fixed to 5.
    \item Dark matter fraction [\fdm].
    \item Concentration of the dark matter halo ($c$) that is fixed to 3.5.
\end{itemize}
and the parameters for the kinematics,
\begin{itemize}
    \item Inclination angle ($i$)
    \item Position angle ($\phi$)
    \item Kinematic center ($x_{\rm cen}, y_{\rm cen}$)
    \item Systematic velocity ($V_{\rm sys}$)
    \item Velocity dispersion ($\sigma$).
\end{itemize}
Among these nine parameters are set to be free during the fitting ($M_{\rm bar},f_\star, R_\star, R_{\rm gas}, f_{\rm DM}(R<R_{\rm e}),i,\phi,(x_{\rm cen}, y_{\rm cen})$, and $\sigma$). The other parameters are fixed to some certain values and the potential effect of these choices will be further discussed in Section \ref{sec5.2: prior}.

We conducted 3-dimensional fitting of the \dysmalpy\ model on both high- and low-resolution data sets simultaneously, which is a key feature implemented in the latest version of this software. We utilize a masked region as described in Section \ref{sec3.2: bbarolo}. Initially, we ran the MCMC sampling process with 150 walkers, each taking 800 steps, following a burn-in phase of 500 steps. After the burn-in phase, we identified the ``maximum a posteriori'' (MAP) values for the parameters. These MAP values were then used as the parameter starting values for the second round of MCMC sampling, using the same number of walkers and steps. The resulting chains, consisting of $150 \times 800$ points, were used to construct the full posterior probability distribution (PDF). We used the 50th percentile of the MCMC samples as the best-fit results and calculated the 16th and 84th percentiles of the MCMC samples to determine the uncertainties of the parameters.

\subsection{\bbarolo}
\label{sec3.2: bbarolo}

The non-parametric, \bbarolo\footnote{\url{https://bbarolo.readthedocs.io/en/latest/}} \citep{DiTeodoro&Fraternali2015} is a tilted-ring model that separates a rotating gaseous disk into several concentric rings \citep{Rogstad+1974}. It was developed for a wide range of applications to emission-line data cubes, and has been widely used in analysing gas kinematics of galaxies at both low-$z$ \citep{Iorio+2017, ManeraPina+2019, Tan+2019, BewketuBelete+2021, Deg+2022, Perna+2022, Fei+2023, Wong+2024} and high-$z$ \citep{DiTeodoro+2016, Loiacono+2019, Bischetti+2021, Fraternali+2021, Fujimoto+2021, Fujimoto+2024, Jones+2021, Sharma+2021, Sharma+2022, Sharma+2023, Lelli+2023, Pope+2023, Posses+2023, Rizzo+2023, RomanOliveira+2023}.

In particular, \bbarolo\ firstly creates a physical hypercube model consisting of concentric rings. These rings are defined by their radius, width, spatial center, systemic velocity ($V_{\rm sys}$), inclination ($i$), position angle ($\phi$), rotational velocity $V_{\rm rot}$, velocity dispersion $\sigma$, face-on gas surface density ($\Sigma$), and scale-height $z0$. It then populates this space with clouds within a ring-shaped space by drawing from a set of input parameters. This is repeated for all rings. The physical cube is then collapsed along the line-of-sight to create a mock data cube, whose three axes are R.A., Dec., and los velocity, for comparison with observations. \citet{DiTeodoro&Fraternali2015} provide extensive details and tests of the impact of spatial and spectral resolution, inclination, and signal-to-noise ratio (S/N) using data from local galaxies and mock models.

The non-parametric model commonly requires a high SNR to separate the signal from the noise. Therefore, we used \bbarolo\ to create a mask region for two data sets before performing the fitting procedure. 
For our sample, we selected the parameters \texttt{SNRCUT} and \texttt{GROWTHCUT} to define the primary and secondary SNR cuts. The algorithm identifies pixels with flux above a threshold defined by \texttt{SNRCUT}, and then increases the detection area by adding nearby pixels that are above a secondary threshold defined by \texttt{GROWTHCUT}. To ensure that we captured real signal and removed bad pixels, we used \texttt{SNRCUT} values of 5 and \texttt{GROWTHCUT} of 3 for the high-resolution data. For the low-resolution data, we set the \texttt{SNRCUT} and \texttt{GROWTHCUT} to be 4 and 2.5, respectively. Then we model the high- and low-resolution data separately. The final models are normalized to the zeroth-moment map of the observations pixel-by-pixel.

We then obtained the rotation velocities and velocity dispersions from fitting with \bbarolo\ while applying the mask described above. To produce smooth RCs, we utilized the two-stage fitting mode provided by \bbarolo, where parameters are regularized using specific functions during the second stage. We modified the two-step procedure outlined in \cite{Fei+2023} to fit the gas kinematics. During the first stage of fitting, we allowed the kinematic center, systematic velocity, rotation velocity, velocity dispersion, position angle, and inclination angle to be free parameters. The ring width was set to half of the FWHM of the beam size in the fitting process. We note that for both high- and low-resolution data, the \cii\ maps are resolved into 4 to 6 spatial elements, sufficient for kinematic modeling with \bbarolo \citep{DiTeodoro&Fraternali2015}. The initial estimates for the systemic velocity, kinematic center, position angles, and inclination angles were set to the best-fit results obtained from Sec.~\ref{sec3.1: dysmalpy}, specific for our targets. In the second stage, we employed regularization with a constant function for the kinematic center, the systemic velocity, the inclination, and the position angle, resulting in only two free parameters, rotation velocity and velocity dispersion. 

As a summary, \bbarolo\ has the following parameters,
\begin{itemize}
    \item Center of the galaxy ($x_0,y_0$)
    \item Systematic velocity ($V_{\rm sys}$)
    \item Rotation velocity of each ring ($V_{\rm rot}$)
    \item Velocity dispersion of the ring ($\sigma$)
    \item Position angle of the kinematic major axis ($\phi$)
    \item Inclination angle ($i$)
\end{itemize}
where $V_{\rm rot}, \sigma, \phi,$ and $i$ are free parameters. The center of galaxies ($x_0,y_0,V_{\rm sys}$) are fixed to the best-fit value from \dysmalpy\ fitting.

\begin{figure*}
    \centering
    \includegraphics[width=\linewidth]{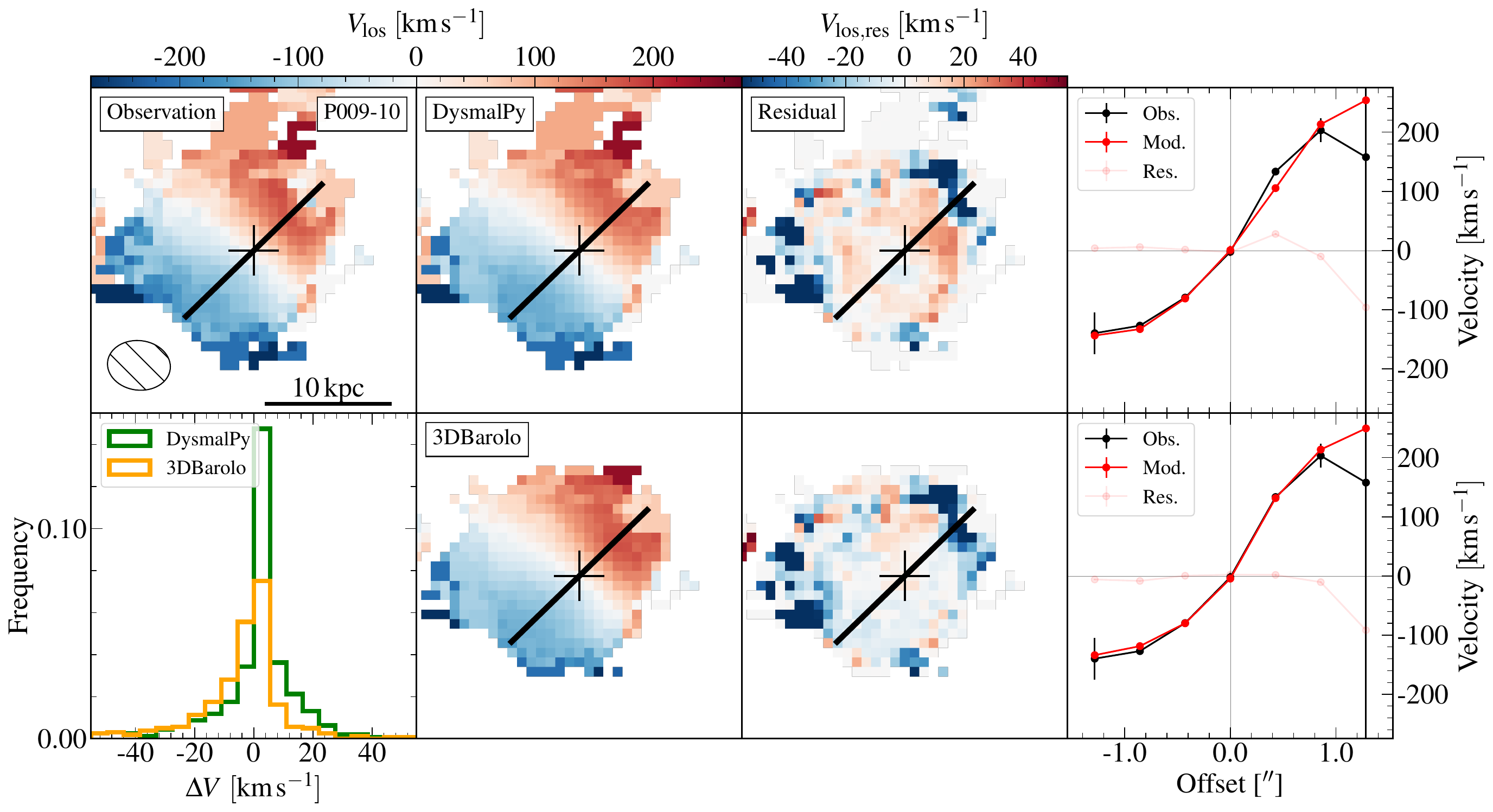}
    \includegraphics[width=\linewidth]{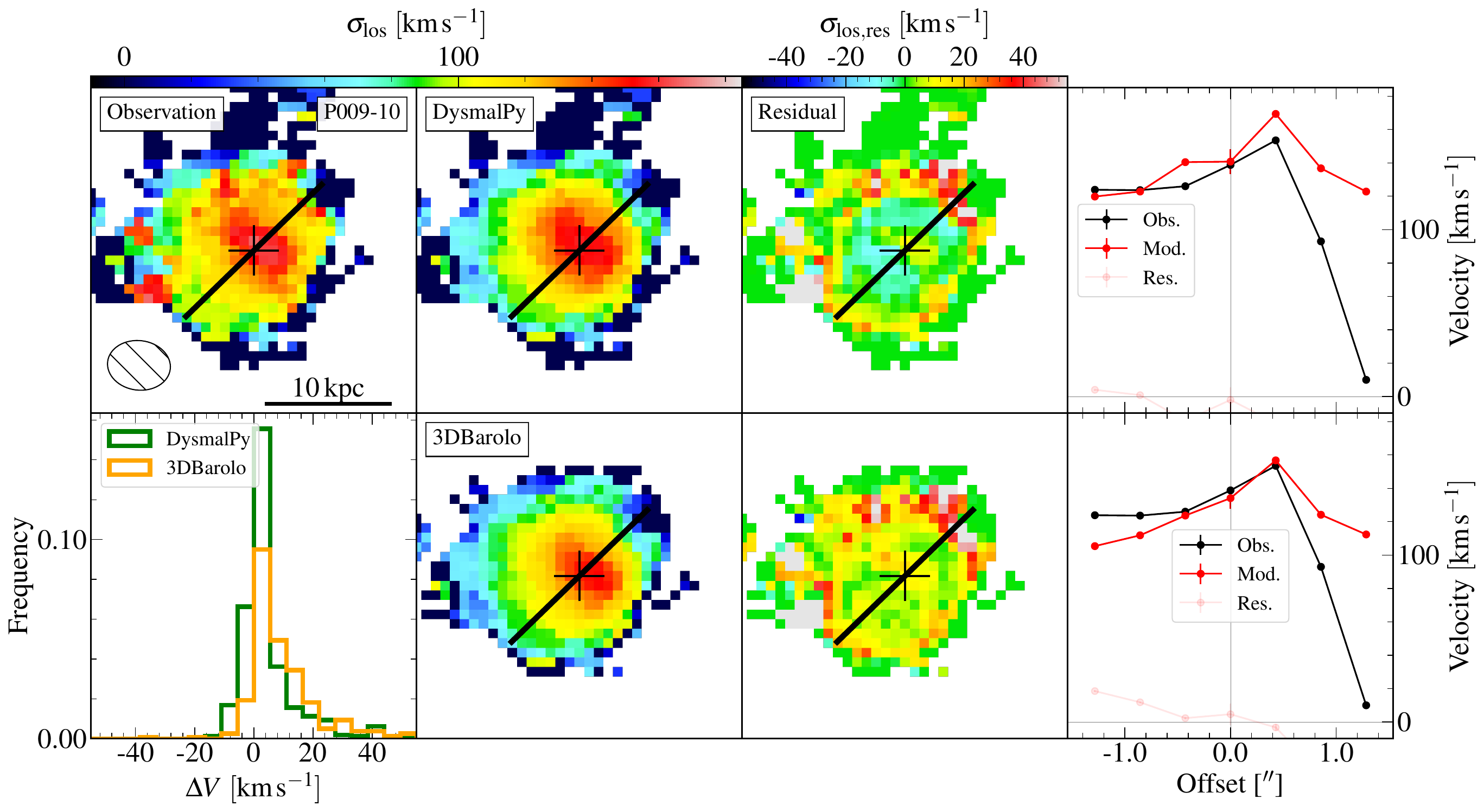}
    \caption{Observed and model \cii\ los-velocity (top) and dispersion (bottom) maps (low-resolution data) of \targb. In each sub-panel, the first row presents the observed \cii\ moment maps (column 1) with the best-fit kinematic model (column 2), with the top row showing the model from \dysmalpy~ and the bottom row showing the \bbarolo\ results. The third column displays the residual from the fit to the velocity map for the corresponding kinematic model. The fourth column shows the 1D velocity profile across the kinematic major axis for the data, model, and residual. The bottom left panel presents the 1D histogram of the residual between observation and model, respectively.}
    \label{Fig3: targb}
\end{figure*}

\begin{figure*}
    \centering
    \includegraphics[width=\linewidth]{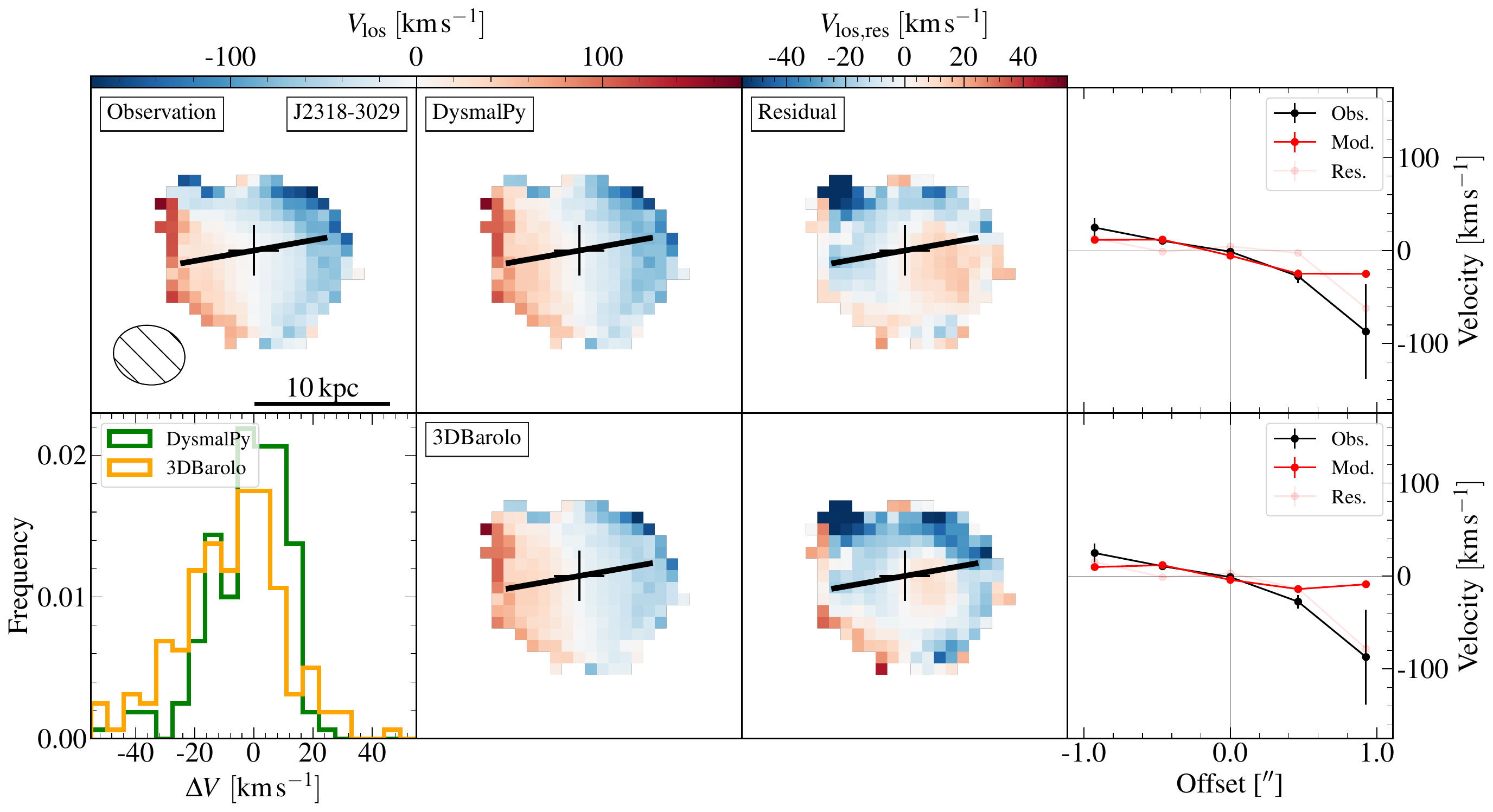}
    \includegraphics[width=\linewidth]{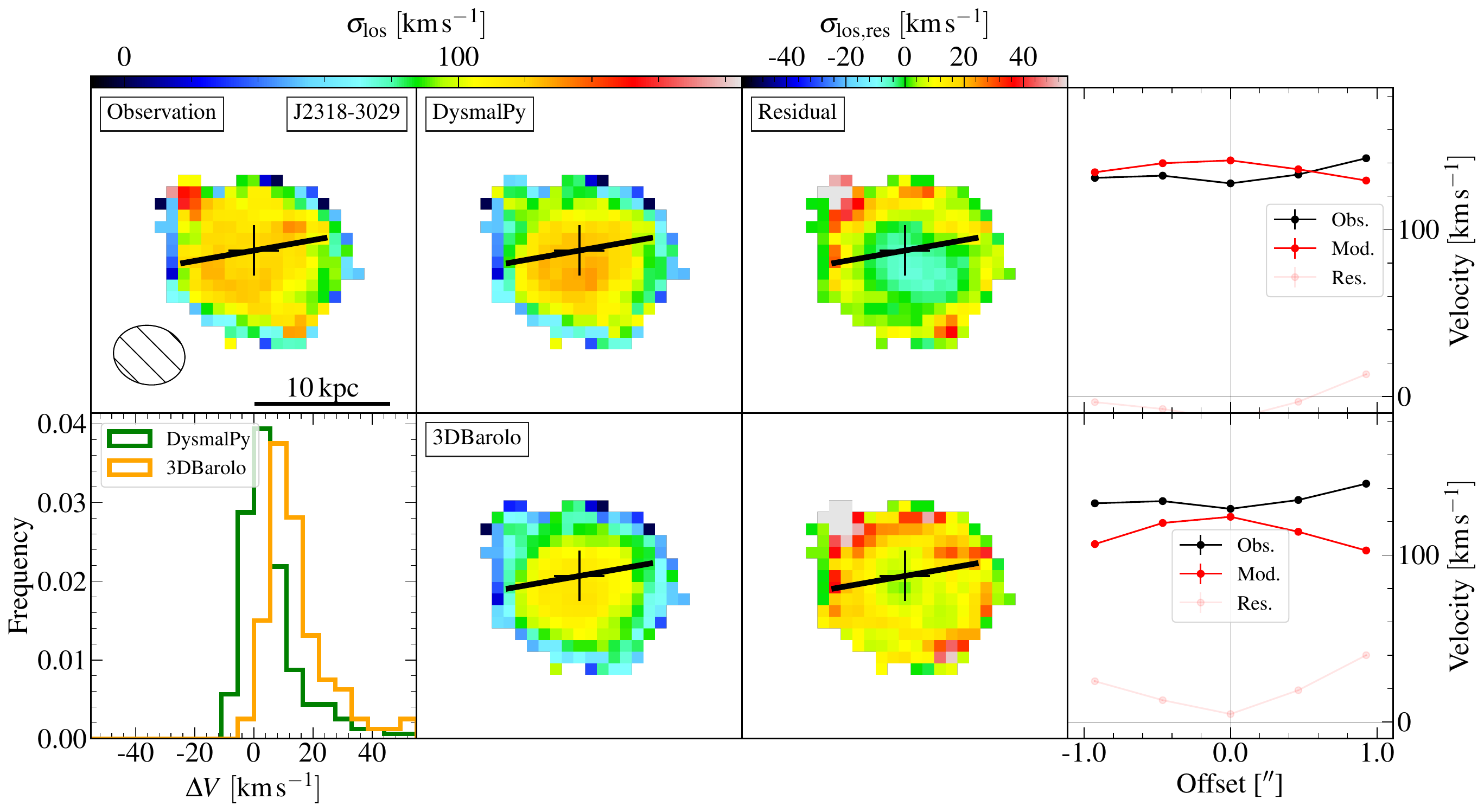}
    \caption{Similar to the Figure \ref{Fig3: targb}, but for the target \targa.}
    \label{Fig4: targa}
\end{figure*}

\section{Results}
\label{sec4: results}

\subsection{Spatial distribution of  \cii\ emission}
\label{sec4.1: cii distribution}

For \targb, prominent \cii\ emission extends well beyond  $r\sim 10\,\rm kpc$ (Figure~\ref{Fig1: intensity map}). The \cii\ distribution (3$\sigma$ contour) has an extension to the north, east, and west from the quasar. This extended emission likely has a more complex origin, possibly involving quasar feedback, gas recycling, and/or gas stripped from potential companions/satellites. Despite this complexity, we isolated a regularly rotating component for our analysis (Sec.~\ref{sec3.3: removing non-circular gas}).

Regarding \targa, the \cii\ distribution appears to be more symmetric. 
The analysis of the $uv$-visibility of this target indicates the detection of the extended \cii\ emission being fairly compact. After performing CASA \texttt{imfit} to the high- and low-resolution intensity maps, we find that the \cii\ flux derived from high-resolution data is $1.50\,\rm Jy\,km\,s^{-1}$, smaller than that derived from the low-resolution data, which is $1.70\,\rm Jy\,km\,s^{-1}$. The analysis indicates that there is 13\,\% additional extended emission which is resolved out with the high-resolution observations.

\begin{deluxetable*}{ccccccccccc}
    \tablenum{3}
    \tablecaption{\dysmalpy\ fitting results \label{Tab3: dysmal}}
    \tablewidth{0pt}
    \tablehead{
    \colhead{Name} & \colhead{$\log M_b$} & \colhead{$f_\star$} & \colhead{$R_\star$} & \colhead{$n_\star$} & \colhead{$R_{\rm gas}$} & {$n_{\rm gas}$} & \colhead{\fdm} & \colhead{$\sigma_{\rm low}$} & \colhead{$\sigma_{\rm high}$} & \colhead{$\log M_{\rm halo}$} \\ 
    \colhead{} & \colhead{($M_\odot$)} & \colhead{} & \colhead{} & \colhead{} & \colhead{} & \colhead{} & \colhead{(kpc)} & \colhead{($\rm km\,s^{-1}$)} & \colhead{($\rm km\,s^{-1}$)} & \colhead{($M_\odot$)}  
    }
    \decimalcolnumbers
    \startdata
        Prior & [9, 12] & [0, 1] & [0, 3] & Fixed & [0.1, 5] & Fixed & [0, 1] & [5, 200] & [5, 200] & Tied \\
        \targb & $10.84_{-0.13}^{+0.12}$ & $0.50_{-0.17}^{+0.15}$ & $1.96_{-0.44}^{+0.41}$ & $1.0$ & $2.77_{-0.75}^{+0.51}$ & 1.36 & $0.61_{-0.08}^{+0.08}$ & $115.85_{-5.32}^{+5.99}$ & $125.23_{-9.52}^{+10.57}$ & $12.85_{-0.21}^{+0.20}$ \\ 
        \targa\ & $10.72_{-0.29}^{+0.22}$ & $0.54_{-0.21}^{+0.21}$ & $1.30_{-0.45}^{+0.76}$ & $1.0$ & $2.57_{-0.86}^{+1.13}$ & 2.78 & $0.53_{-0.23}^{+0.20}$ & $120.37_{-4.70}^{+7.67}$ & $111.04_{-5.94}^{+16.12}$ & $12.50_{-0.84}^{+0.66}$\\
    \enddata
    \tablecomments{(1) Name of targets; (2) The total baryonic mass of the galaxy; (3) The ratio between the stellar mass (refer to bulge mass according to our assumption) and the total baryon mass; (4) The effective radius of the stellar component; (5) The S\'ersic index of the stellar component; (6) The effective radius of the gas component; (7) The S\'ersic index of the gas component, fixed to the fitting result of the \cii\ intensity map; (8) The dark matter mass fraction within the effective radius ($R_e$), where $R_e$ denotes the half mass radius of the gas component; (9) The velocity dispersion of the low-resolution data; (10) The velocity dispersion of the high-resolution data; (11) The virial mass of the dark matter halo, estimated from the combination of dark matter fraction and the baryonic mass.}
\end{deluxetable*}

\begin{deluxetable*}{ccccccc}
    \tablenum{4}
    \tablecaption{\bbarolo\ fitting results \label{Tab4: barolo}}
    \tablewidth{0pt}
    \tablehead{
    \colhead{Name} & \colhead{Radius} & \colhead{$V(R)$} & \colhead{$\sigma(R)$} & \colhead{$i$} & \colhead{$\phi$} & \colhead{$V/\sigma$} \\
    \colhead{} & \colhead{(kpc)} & \colhead{($\rm km\,s^{-1}$)} & \colhead{($\rm km\,s^{-1}$)} & \colhead{($^\circ$)} & \colhead{($^\circ$)} & \colhead{} 
    }
    \decimalcolnumbers
    \startdata
        Boundary$^1$ & & $[0, 500]$ & $[5, 200]$ & $[0, 180]$ & $[-180, 180]$ & \\
        \targb(high-res)  & 0.29 & $128_{-25}^{+26}$ & $135_{-9}^{+9}$ & $40\pm 12$ & $-48\pm11$ & $0.9\pm0.3$ \\
        & 0.88 & $228_{-22}^{+18}$ & $97_{-9}^{+9}$ & $40\pm 12$ & $-48\pm11$ & $2.4\pm0.4$ \\
        & 1.46 & $235_{-25}^{+28}$ & $106_{-12}^{+14}$ & $40\pm 12$ & $-48\pm11$ &  $2.2\pm0.5$ \\
        \targb(low-res) & 0.96 & $260_{-10}^{+14}$ & $86_{-6}^{+6}$ & $37\pm 5$ & $-47\pm10$ & $3.0\pm0.4$ \\
        & 2.89 & $287_{-13}^{+10}$ & $106_{-5}^{+6}$ & $37\pm 5$ & $-47\pm10$ &  $2.7\pm0.3$ \\
        & 4.82 & $310_{-12}^{+13}$ & $111_{-6}^{+6}$ & $37\pm 5$ & $-47\pm10$ &  $2.8\pm0.3$ \\
        & 6.75 & $325_{-10}^{+13}$ & $100_{-5}^{+7}$ & $37\pm 5$ & $-47\pm10$ &  $3.2\pm0.3$ \\
        & 8.67 & $351_{-39}^{+43}$ & $85_{-12}^{+12}$ & $37\pm 5$ & $-47\pm10$ &  $4.1\pm1.1$ \\
        \targa(high-res) & 0.26 & $171_{-12}^{+12}$ & $86_{-3}^{+4}$ & $23\pm5$ & $95\pm10$ &  $2.0\pm0.2$ \\
        & 0.78 & $200_{-27}^{+31}$ & $84_{-6}^{+6}$ & $23\pm5$ & $95\pm10$ &  $2.4\pm0.6$ \\
        \targa(low-res) & 1.30 & $254_{-28}^{+21}$ & $111_{-3}^{+3}$ & $18\pm5$ & $93\pm10$ &  $2.3\pm0.3$ \\
        & 3.89 & $231_{-36}^{+28}$ & $114_{-4}^{+5}$ & $18\pm5$ & $93\pm10$ &  $2.0\pm0.4$ \\
        & 6.49 & $146_{-33}^{+33}$ & $78_{-9}^{+10}$ & $18\pm5$ & $93\pm10$ &  $1.9\pm0.7$ \\
    \enddata
    \tablecomments{(1) Name of the targets; (2) Radius of the center of each ring; (3) The rotation velocity of the corresponding ring; (4) The velocity dispersion; (5) The inclination angle used in the 2nd iteration of \bbarolo\ fitting, with uncertainties derived from the 1st step; (6) The position angle used in the 2nd iteration of \bbarolo\ fitting, with uncertainties derived from the 1st step; (7) The ratio between the rotation velocity and velocity dispersion for each concentric ring.\\
    $^1$ The boundary of fitting parameters for \bbarolo.}
\end{deluxetable*}

To quantitatively describe the distribution of the \cii\ emission observed in the new datasets, we conducted S\'ersic model fitting for the low-resolution data. The model was created using the \texttt{Astropy} \citep{Astropy+2013} and then convolved with the synthesized beam to match the observations using the \texttt{convolve} task in \texttt{Astropy}. We took into account the statistical uncertainties introduced by the correlated noise and performed the fitting following the method outlined in \citet{Tsukui+2023}. The best-fit result indicates an effective radius of $3.10\pm0.12$ and $1.21\pm0.06$\,kpc for \targb\ and \targa, with S\'ersic indices of $1.36_{-0.11}^{+0.12}$ and $2.79_{-0.57}^{+0.75}$, respectively. 

\subsection{Kinematic modeling}
\label{sec4.2: kinematics}

We present the best-fit kinematic models to the observations using two fitting codes in Figure \ref{Fig3: targb} and \ref{Fig4: targa}. The parameters estimated from \dysmalpy\ are listed in Table \ref{Tab3: dysmal}. The velocity maps for the observation and the model produced by two codes are generated with the same mask (Sec. \ref{sec3.2: bbarolo}). We note that for both targets, the majority of the residual is within the range of $-30$ to $30\,\rm km\,s^{-1}$ which is probably attributed to the limited velocity resolution, suggesting that our targets can be well described by our model. 

\begin{figure*}
    \centering
    \includegraphics[width=0.49\linewidth]{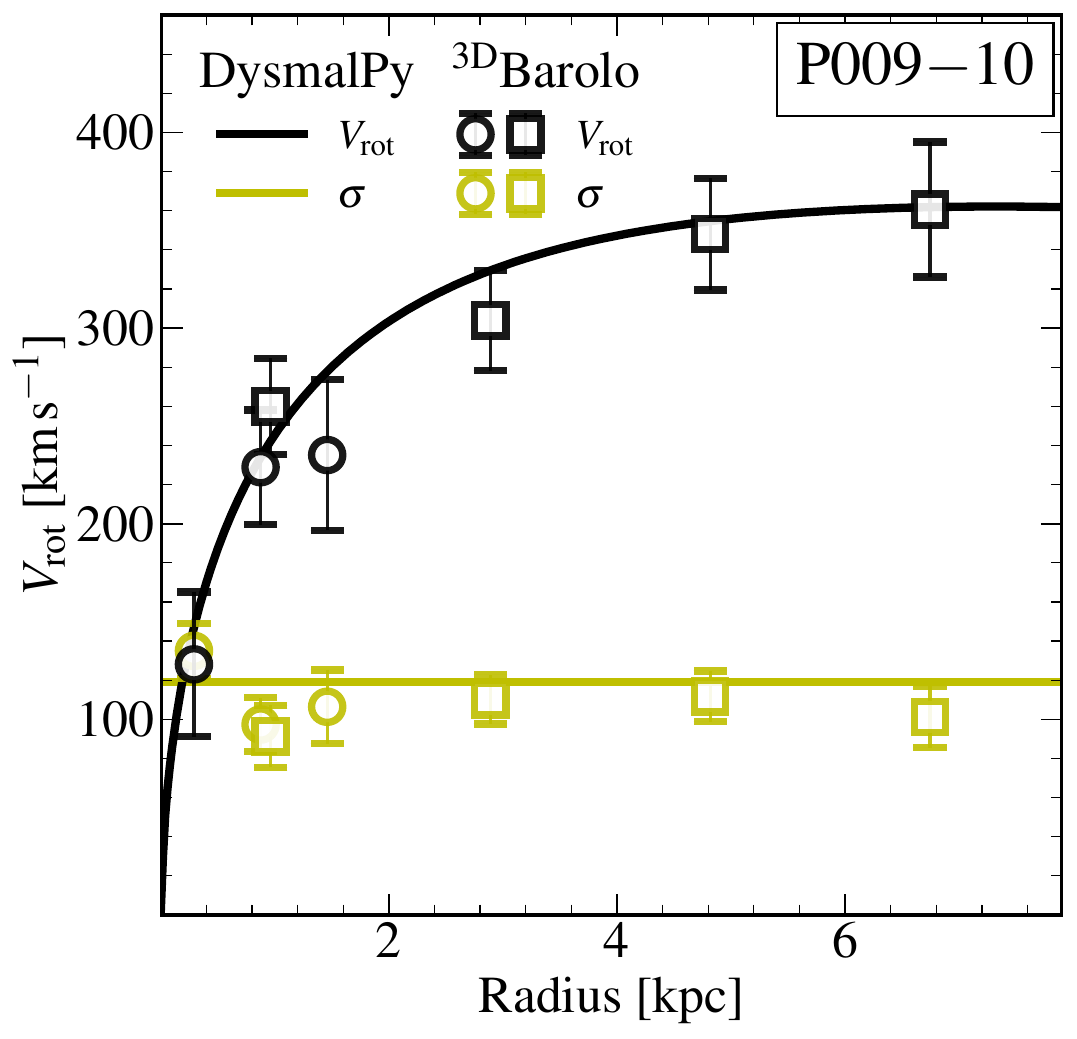}
    \includegraphics[width=0.49\linewidth]{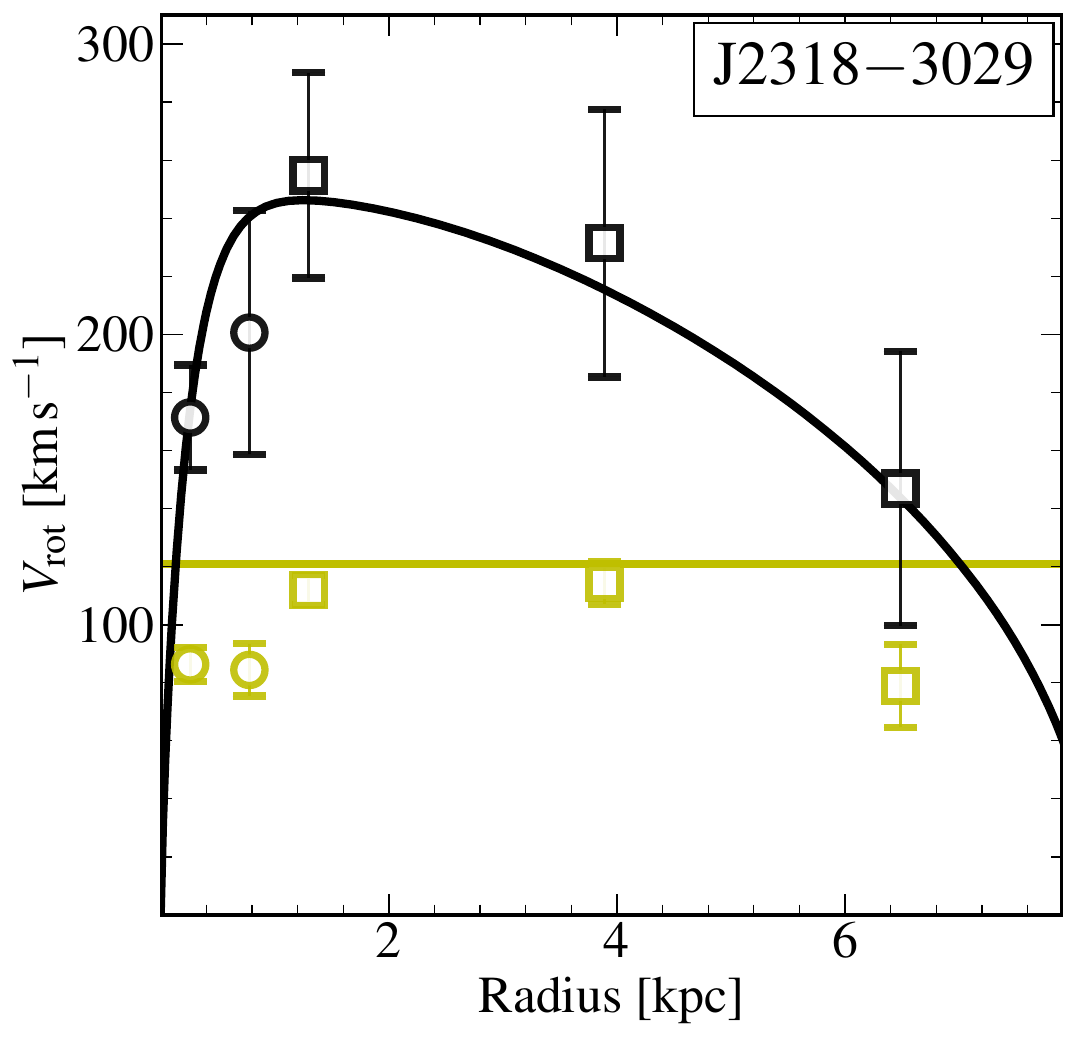}
    \includegraphics[width=0.49\linewidth]{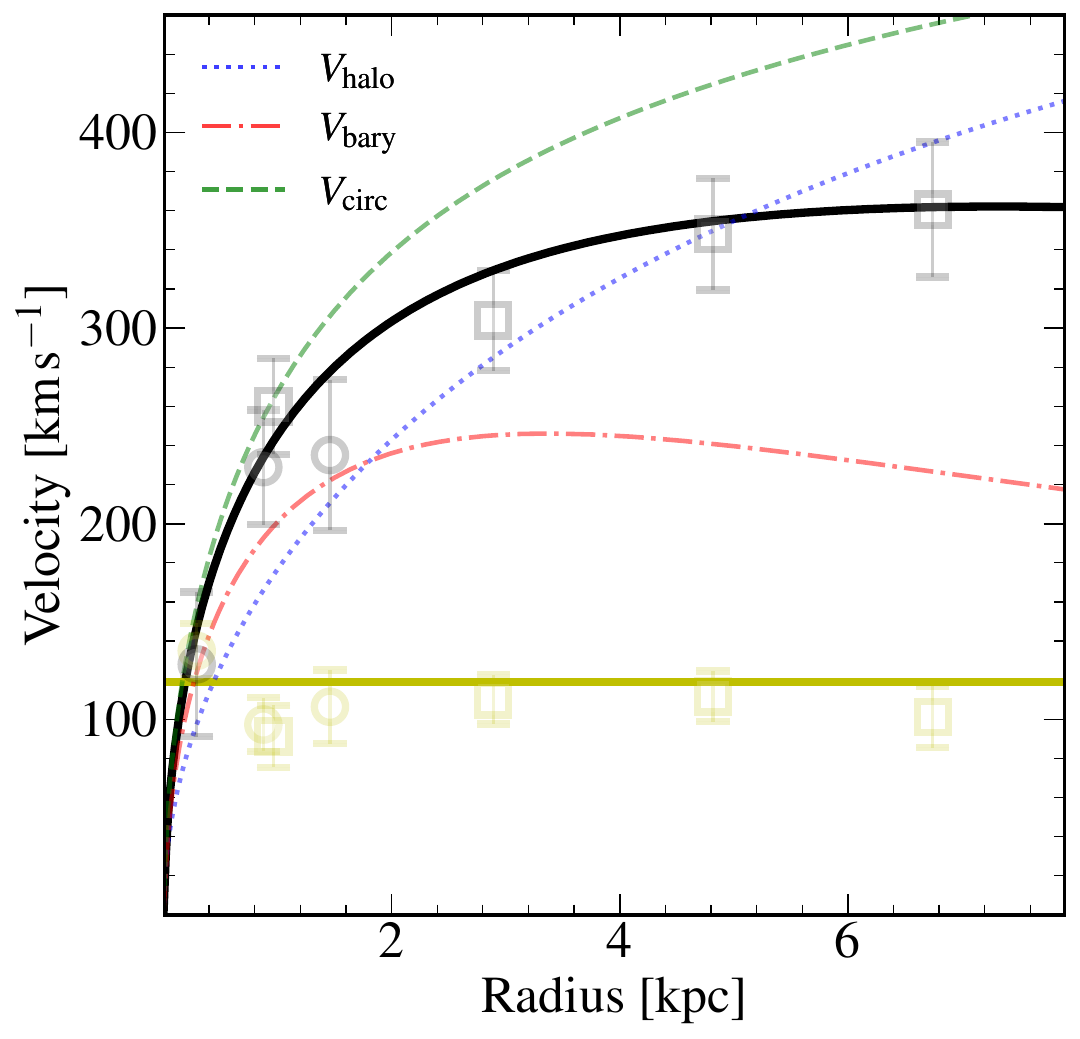}
    \includegraphics[width=0.49\linewidth]{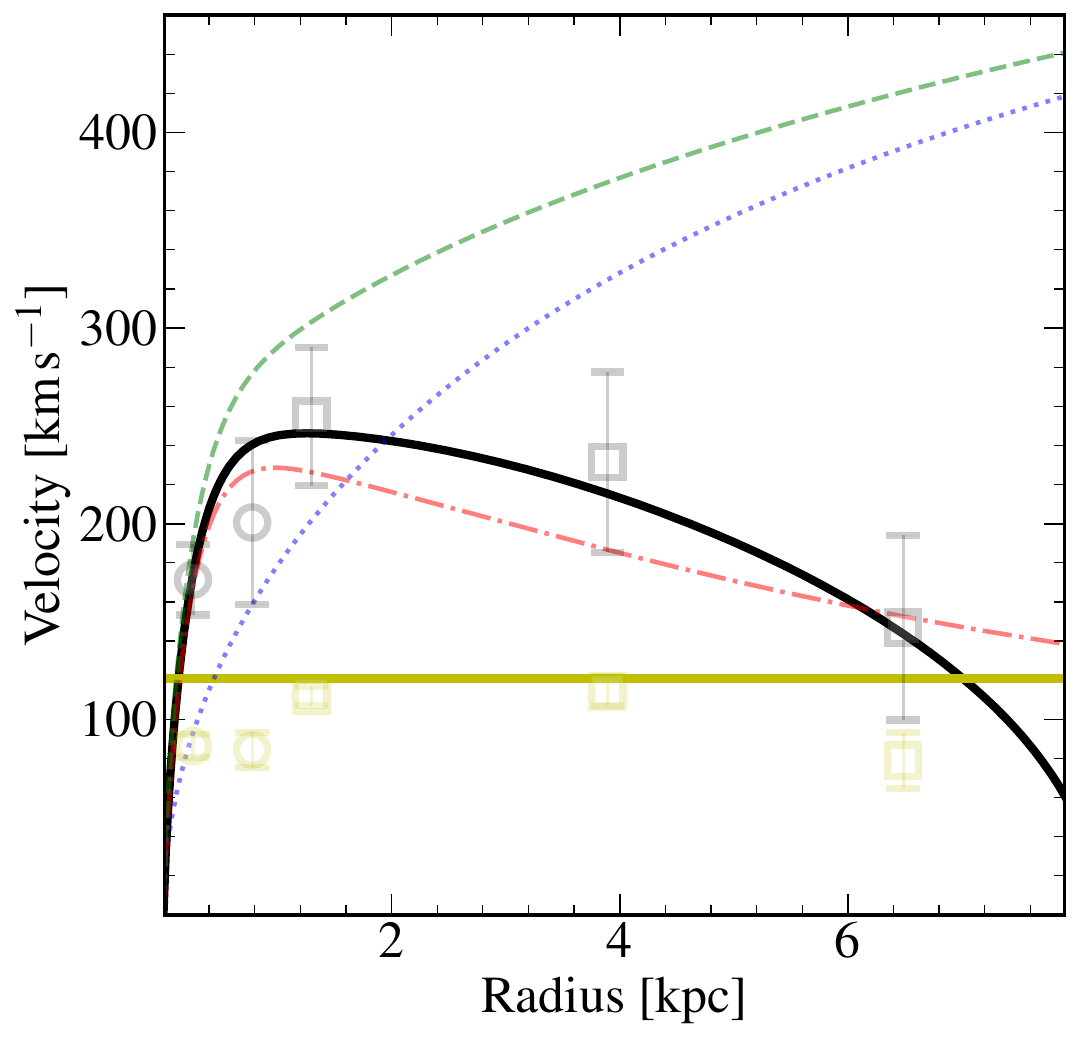}    
    \caption{Rotation curves for \targb~(left column) ~and \targa~(right column). {\it Top row:} Rotation velocity and velocity dispersion derived from the best-fit \dysmalpy\ and \bbarolo\ models. Open circles and squares denote the values from high- and low-resolution data, respectively, and the black and yellow solid line present the rotation velocity and velocity dispersion as a function of radius derived from best-fit \dysmalpy\ model. {\it Bottom row:} 
    Same as above with the addition of the circular velocities (green dashed line) after accounting for the asymmetric drift correction, reflecting the entire gravitational potential without the influence of gas pressure. The red dash-dotted and blue dotted lines illustrate the contributions of baryonic matter and dark matter to the circular velocity, respectively, derived by decomposing the overall circular velocities.}
    \label{fig6: rcs}
\end{figure*}

\subsubsection{\targb}

The velocity field of \targb\ exhibits complex kinematic properties, likely resulting from different components (Fig. \ref{fig2: filtering}a). Upon removing the external component, the remaining \cii\ emission reveals a more clear velocity gradient, that can be fitted with kinematic models with negligible contamination. Both kinematic models successfully replicate the observed line-of-sight velocity and velocity dispersion within the uncertainties (Fig. \ref{Fig3: targb}). However, we observed slight differences in the velocity dispersion profiles generated by the two models (see Tables \ref{Tab3: dysmal} and \ref{Tab4: barolo}), which may be attributable to differences in the models assumptions. Specifically, in \dysmalpy, the velocity dispersion is assumed to be constant across the radius, whereas in \bbarolo, the velocity dispersion of each ring can vary independently. Nonetheless, the velocity dispersion profiles generated by both models are consistent with each other and with the observational data (Fig.~\ref{fig5: rc_comp}). 

\subsubsection{\targa}

The velocity map of \targa\ presents a clear gradient, with the blue-shifted component on the west side and the red-shifted component on the east (Fig.~\ref{Fig4: targa}). This velocity gradient is successfully reproduced by \dysmalpy\ and \bbarolo. A larger residual can be seen in the histogram of the residual of \bbarolo, which shows that the residual of \bbarolo\ spans a broader range compared to the result obtained by \dysmalpy. This means that for this target, \dysmalpy\ provides a better description of the gas kinematics. The residual velocity is smaller than 40\,$\rm km\,s^{-1}$ which can be attributed to the limited velocity resolution (Table \ref{Tab2: cubes}), suggesting the gas motion in this target well matches a rotating disk. Additionally, the velocity residual becomes larger at the edge of the galaxy ($R\gtrsim 1\farcs5$), probably due to the relatively low SNR.

The lower panel of Figure \ref{Fig4: targa} displays the observed velocity dispersion map alongside the best-fit models. The observed map reveals a nearly constant velocity dispersion of approximately $\sigma \sim 100\,\rm km\,s^{-1}$ across most radii, with a noticeable decline at the outskirts of the galaxy, potentially due to the relatively lower SNR in this region \citep{Fujimoto+2021}. The velocity dispersion maps generated by the \dysmalpy\ and \bbarolo\ models exhibit a similar pattern. We observed a slight difference between the velocity dispersion maps produced by the two models, though they remain consistent within the uncertainties. 

\subsection{Rotation curves for quasar host galaxies at $z\sim 6$}
\label{4.3: rcs}
\begin{figure*}
    \centering
    \includegraphics[width=\linewidth]{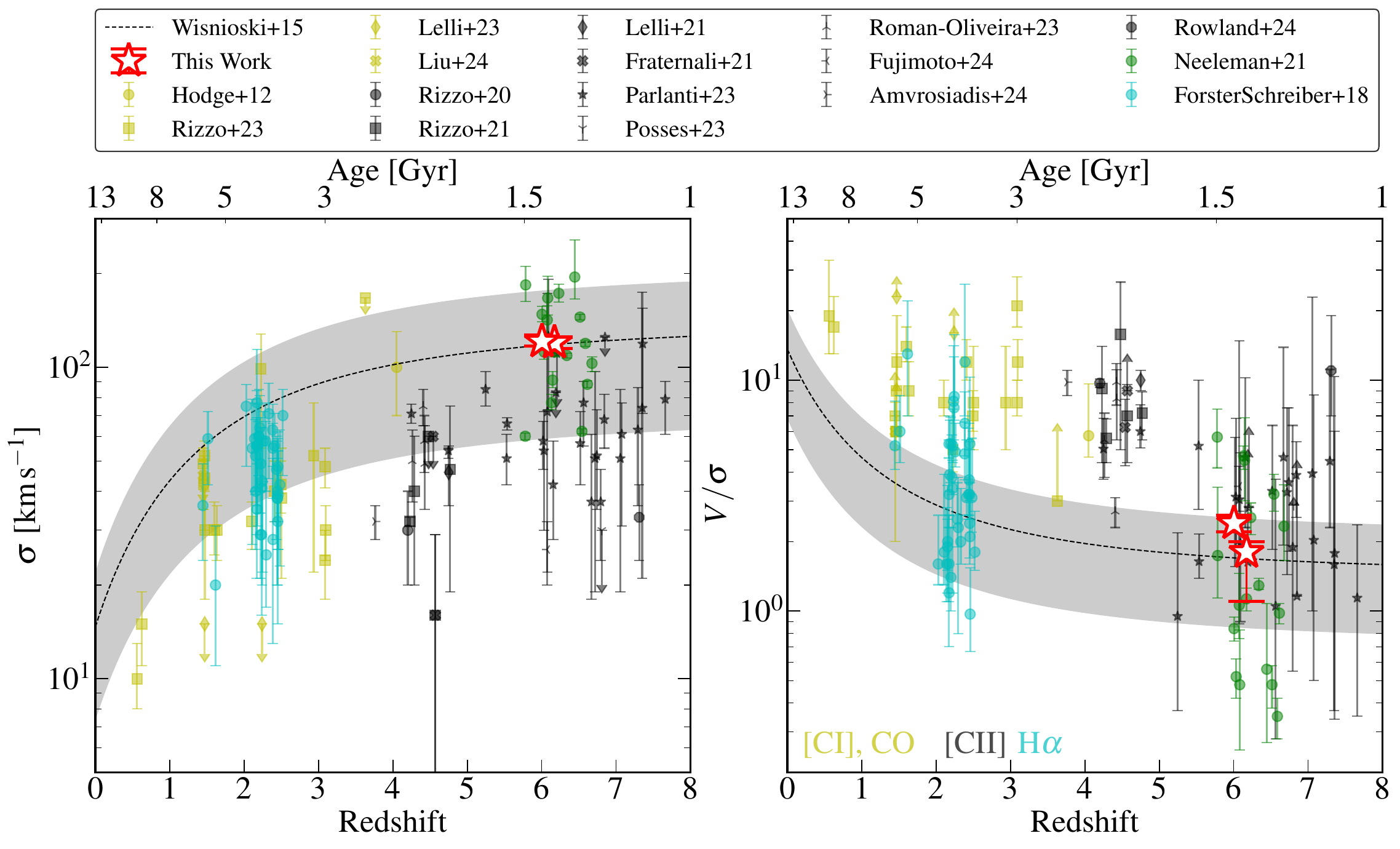}
    \caption{{\it Left panel:} Velocity dispersion versus redshift for our targets (red stars) and galaxies from the literature with observations of CO or \ci~\citep[yellow;][]{Hodge+2012, Rizzo+2023, Lelli+2023, Liu+2024}, \cii~\citep[gray;][]{Rizzo+2020, Rizzo+2021, Lelli+2021, Fraternali+2021, Parlanti+2023, Posses+2023, RomanOliveira+2023, Fujimoto+2024, Amvrosiadis+2024, Rowland+2024}, and \Ha~\citep[cyan;][]{ForsterSchreiber+2018}. Quasars at the same redshifts are shown as green points \citep{Neeleman+2021}. The gray dashed line illustrates the expected velocity dispersion as a function of redshift for galaxies with $\log M_\star/M_\odot=10.5$ and $V_{\rm rot}=200\,\rm km\,s^{-1}$, according to the semi-analytic model \citep{Wisnioski+2015}. The gray-shaded region denotes expected velocity dispersion for galaxies with same mass, while for $100<V_{\rm rot}<300\,\rm km\,s^{-1}$, respectively. {\it Right panel:} Ratio between the regular rotation and the random motion ($V/\sigma$) as a function of redshift. The gray dashed line represents the expected $V/\sigma$ of a quasi-stable (Toomre $Q=1.0$) disk within a galaxy with $\log \left(M_\star/M_\odot\right)=10.5$ as a function of redshift. The gray-shaded region denotes the expected $V/\sigma$ with $0.67<Q<2$, respectively \citep{Wisnioski+2015}.}
    \label{fig7: v/sigma}
\end{figure*}

We now determine the kinematics and dynamics of the \cii-emitting gas out to the extended regions of our two quasars at $z\sim 6$. By combining the low-resolution data with previous high-resolution data, we generate the RCs, as shown in Figure \ref{fig6: rcs}, which depicts the intrinsic rotation velocities after correcting for observational effects (e.g., inclination, beam-smearing) from the inner region to the outskirts of the galaxy ($r\gtrsim 3-6\,R_e$). After setting a similar inclination angle in two fitting procedures, the rotation velocities derived from \dysmalpy\ closely match those obtained from \bbarolo. 

We find that the rotation velocity of \targb\ increases steadily and gradually as a function of radius, resembling galaxies in the local universe. In contrast, we see in \targa, the rotation velocity has a peak velocity of $\sim 230\,$\kms\ at a distance of approximately $\sim 1\,$kpc. Subsequently, it decreases gradually as we move towards the outskirts of the galaxy, ultimately reaching 0\,\kms\ at $r\gtrsim 7\,$kpc. This characteristic behavior is consistent with a truncated gaseous disk model where pressure-support results in an asymmetric drift correction at the galaxy outskirts \citep{Genzel+2020}. After accounting for the gas pressure, the circular velocities, seen in the bottom row of Figure~\ref{fig6: rcs}, continue to increase out to $\sim8$ kpc, a likely signature of the presence of a substantial amount of dark matter in their outskirts. Such differences between the circular and rotational velocities are expected for galaxies at high-$z$ with a relatively large velocity dispersion \citep{Genzel+2020}. We note that the effect of gas pressure is slightly smaller in \targb\ rather than in \targa, probably due to the relatively higher $V_{\rm rot}/\sigma$ (Table \ref{Tab4: barolo}). 

\subsection{$V_{\rm rot}/\sigma$ and its evolution}
We examine the ratio between the rotation velocity and the velocity dispersion ($V_{\rm rot}/\sigma$) and our results are shown in Table \ref{Tab4: barolo}. This parameter has been widely used to determine the rotation support of galaxies \citep{Epinat+2009, Burkert+2010, Turner+2017, Fujimoto+2024, Rowland+2024}. Given the truncated feature of \targa, the $V_{\rm rot}/\sigma$ value of this target was averaged from the region of $r_{e, \star}<r<r_{e, g}$ ($r_{e,\star}$ and $r_{e,g}$ are the effective radius of the stellar and gas component, respectively). For \targb, the value of $V_{\rm rot}/\sigma$ does not change a lot after $r>r_{e, g}$. Our results indicate that $V_{\rm rot}/\sigma\sim 2$ for our targets, suggesting our targets are rotation-dominated \citep{Epinat+2009, Newman+2013, Turner+2017}, while dynamically warm disks. 

We then compare the velocity dispersion and the $V_{\rm rot}/\sigma$ ratio with previous studies to explore the evolution of these parameters. Our findings indicate that the velocity dispersion of our targets is relatively higher than that observed in dusty star-forming galaxies (DSFGs) at $z \sim 4-5$, accompanied by a relatively lower $V_{\rm rot}/\sigma$ ratio \citep{Rizzo+2020, Rizzo+2021, Fraternali+2021}. These measurements align with those reported for quasars at similar redshifts (\citealt{Neeleman+2021}; Figure \ref{fig7: v/sigma}). Additionally, our results are consistent with the evolutionary trend predicted by \cite{Wisnioski+2015}, which is driven primarily by the redshift evolution of the gas fraction according to the Toomre stability criterion \citep{Toomre1964}. The concordance between our observations and the predictions from the analytic model suggests a quasi-stable state of cold gas in our targets. 

\begin{figure}
    \centering
    \includegraphics[width=\linewidth]{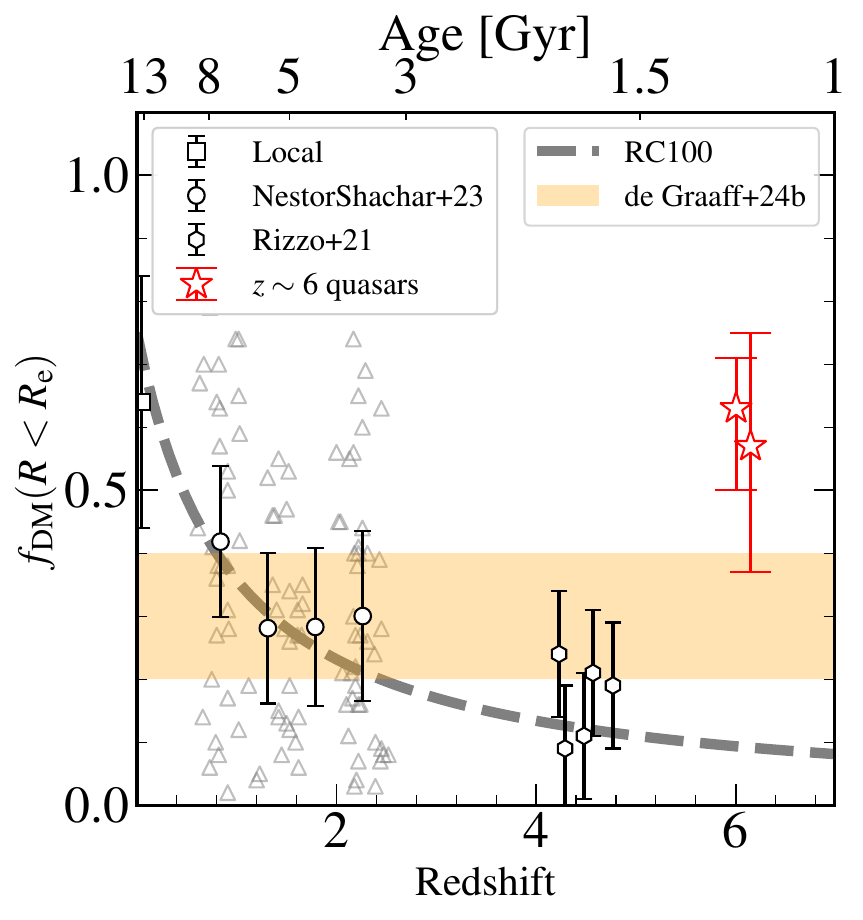}
    \caption{Evolution of \fdm. The \fdm\ of massive star-forming galaxies at $z\sim 2$ are represented as open circles \citep[binned from light gray triangles, ][]{NestorShachar+2023}, and that of several dusty star-forming galaxies at $z\sim 4$ are shown as open hexagons \citep{Rizzo+2021}. The \fdm\ as a function of redshift extrapolated from those previous studies is shown as the grey dashed line. Our results are denoted as red stars, respectively. The orange shaded region represents the typical range of \fdm\ for star-forming galaxies with the stellar mass of $10^{10}-10^{10.5}\,M_\odot$, from simulation perspective \citep{deGraaff+2024b}.}
    \label{Fig7: fdm}
\end{figure}

\subsection{\fdm\ and $M_{\rm h}$}
\label{sec4.4: paras}

The best-fit parameters from \dysmalpy\ are listed in Table \ref{Tab3: dysmal}, and the posterior distribution functions (PDFs) of key parameters, baryonic mass ($\log M_b$), dark matter fraction within the effective radius (\fdm), and dark matter halo mass ($\log M_h$) are shown in Figure \ref{Fig9: corner}. The analysis suggests that the stellar mass of our quasars are on the high-mass end ($\log (M_{\rm b}/M_\odot)\gtrsim 10.5$) compared to galaxies at similar redshift \citep{Stefanon+2021, Morishita+2024}, implying that these SMBHs are hosted by massive galaxies. While our fitting approach does not fully disentangle the degeneracy between stellar and gas components, the best-fit result suggests that the stellar component aligns with the mass-size relation of inactive galaxies at similar redshifts \citep{Morishita+2024}, within the bounds of uncertainty. This finding is consistent with recent detections of stellar light from quasar host galaxies at $z \gtrsim 6$ \citep{Ding+2023, Onoue+2024}. However, given ultra-high-resolution observations, it remains possible that the stellar emission may arise from an ultra-compact structure \citep{Walter+2022, Meyer+2023}. Resolving this degeneracy definitively would require direct observations of the stellar light with JWST.

The dynamic measurements of our targets indicate that they are dark-matter-dominated systems with $f_{\rm DM}(R<R_e) =0.61_{-0.08}^{+0.08}$ and $0.53_{-0.23}^{+0.20}$ (Table \ref{Tab3: dysmal}). When we compare these measurements with the extrapolations from previous studies on massive star-forming galaxies at cosmic noon \citep{NestorShachar+2023},  we find that our values exceed the expected $f_{\rm DM}(R<R_e)\sim 0.2$ by a factor of two (Figure \ref{Fig7: fdm}). These values are also higher than those found in dusty star-forming galaxies at $z\sim 4$ \citep{Rizzo+2021} and the simulation estimation \citep[\fdm~$\sim 0.3$][]{deGraaff+2024b}. The large \fdm~corresponds to dark matter halo masses of $\log (M_h/M_\odot) = {12.85_{-0.21}^{+0.20}}$ and $12.50_{-0.84}^{+0.66}$, respectively, indicating that these quasars reside in the most massive dark matter halos at these redshifts. The massive halo will introduce a galaxy overdensity of $5\sigma$ at these redshifts, which can be examined by the galaxy clustering studies using \jwst. This implies that massive galaxies at $z\sim 2$ are associated with possibly more cored DM halos compared to our targets, indicating that they are at different stages of their evolutionary paths. The substantial dark matter halo mass likely plays a critical role in the evolution of massive galaxies in the early universe. 

\section{Discussion}
\label{sec5: discussion}
\subsection{Comparison between RCs at $z\sim 6$ and at low-$z$}
\begin{figure}
    \centering
    \includegraphics[width=\linewidth]{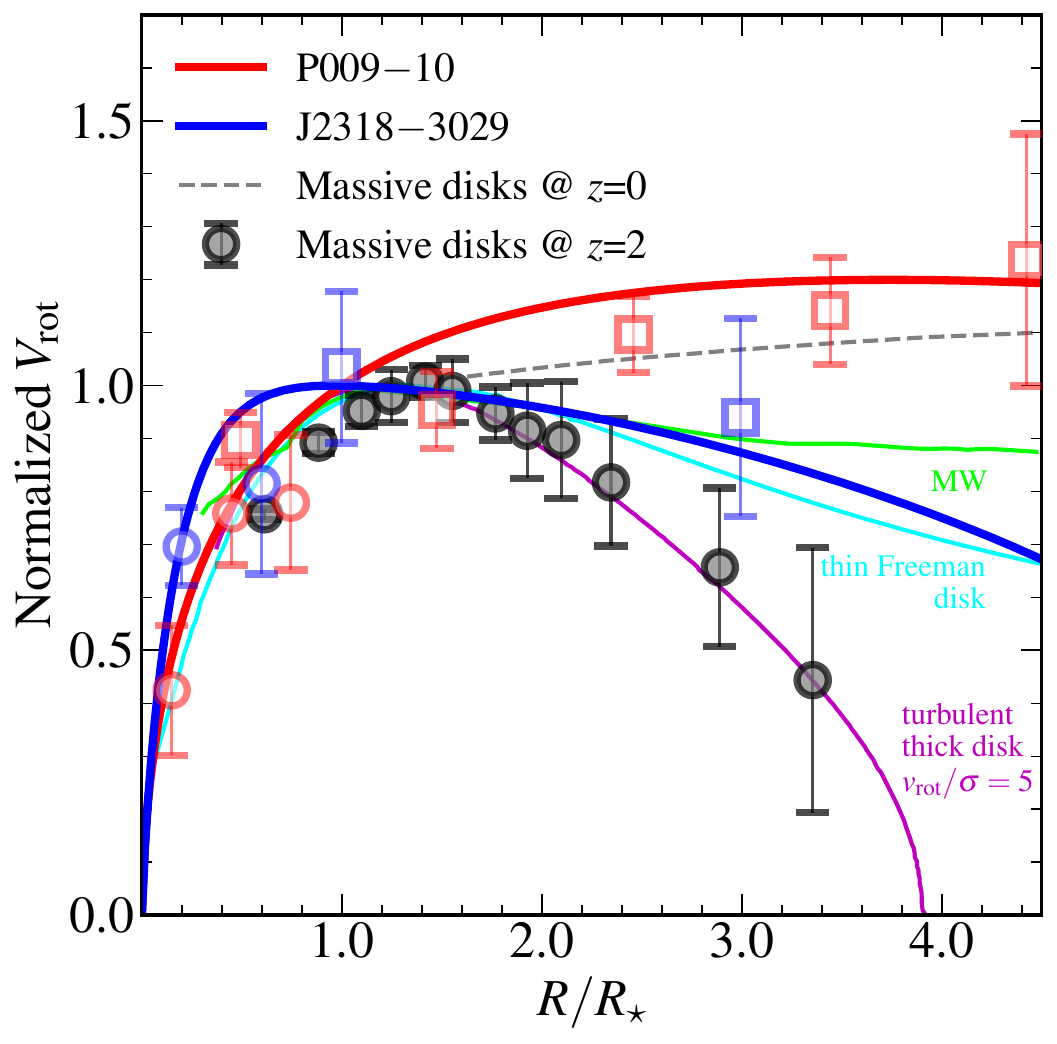}
    \caption{Normalized rotation curves for our targets (red and blue), and comparison with massive star-forming galaxies at relatively lower redshift. The gray data points represent the first six galaxies reported by \cite{Genzel+2017}, and the gray dashed line denotes the averaged rotation curve of the massive star-forming galaxies in the local universe.}
    \label{fig5: rc_comp}
\end{figure}

The extended \cii\ emission enables the measurement of rotation velocities at large radii. Here, we summarize the rotation curve profiles of our targets and compare them with those of massive star-forming galaxies of similar baryon mass ($\sim 10^{10.5}\,M_\odot$) at lower redshifts \citep[e.g.,][]{Genzel+2017}.

Despite the similarity in mass between our targets and the comparison sample, their mass distributions differ significantly due to variations in effective radii, S\'ersic indices, and other structural parameters thus requiring appropriate scaling to compare their rotation curve profiles. To enable a meaningful comparison, we normalized the radii using the effective radius and the RCs using the corresponding $V_{\rm rot}$. The normalized RCs of our targets and the comparison sample are presented in Fig.~\ref{fig5: rc_comp}.

Following their normalization, while the inner slopes of our RCs align closely with those reported in the literature, both \targb\ and \targa\ maintain fairly flattened rotation curves at their outskirts ($R\sim 4\,R_\star$). 
This behavior is consistent with that observed in massive disk galaxies in the local universe but differs from the steep decline seen in an important fraction of massive star-forming galaxies at cosmic noon, where the rotation velocity drops significantly (to nearly 0.1 at $4R_\star$). The pronounced fall-off in the RCs of $z\sim 2$ galaxies is attributed to two main factors: the asymmetric drift correction driven by low $V_{\rm rot}/\sigma$ and a low \fdm. While our targets exhibit relatively lower $V_{\rm rot}/\sigma$, which could lead to a more significant asymmetric drift correction, they possess a comparatively higher \fdm\ than the $z\sim 2$ sample. The increased contribution of dark matter helps to sustain the rotation velocities in the outer regions, resulting in RCs that remain flattened at large radii.

\subsection{Impact of fixing different priors}
\label{sec5.2: prior}
\begin{figure*}
    \centering
    \includegraphics[width=0.45\linewidth]{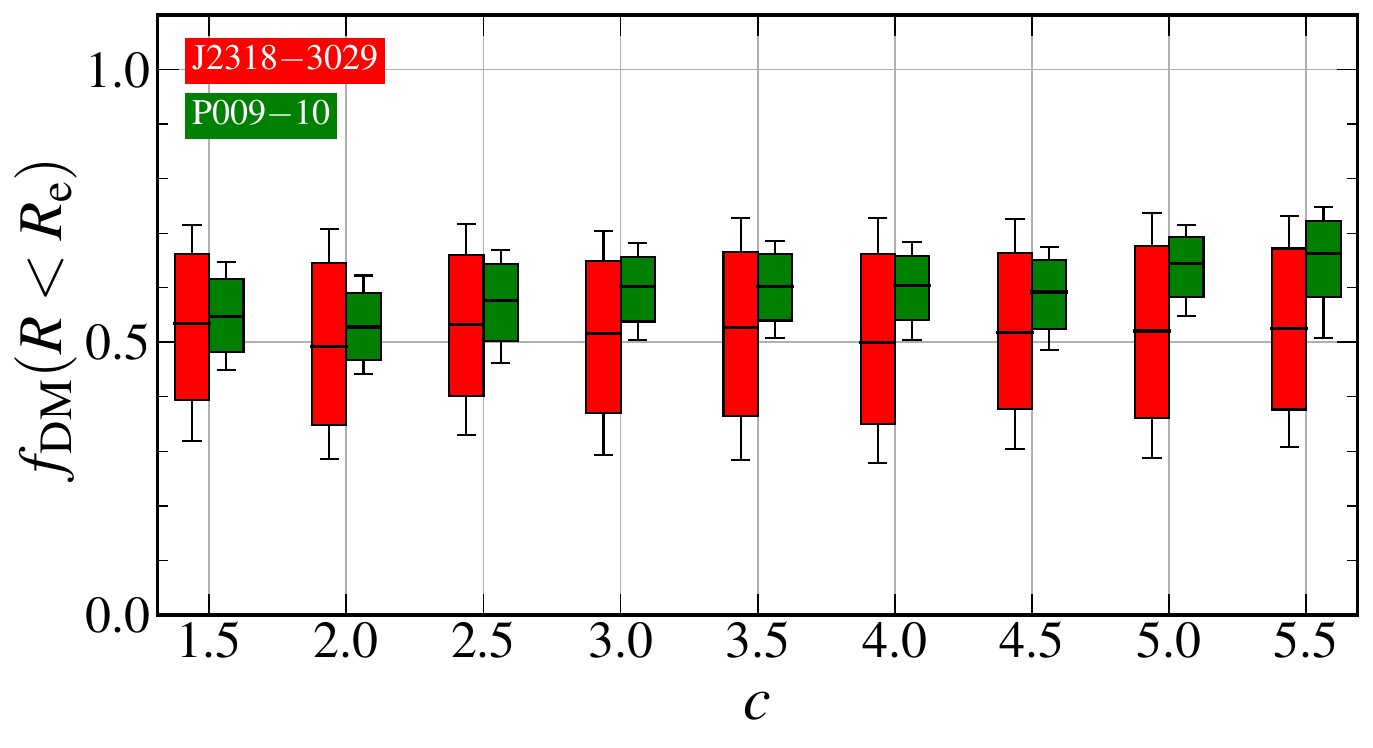}
    \includegraphics[width=0.45\linewidth]{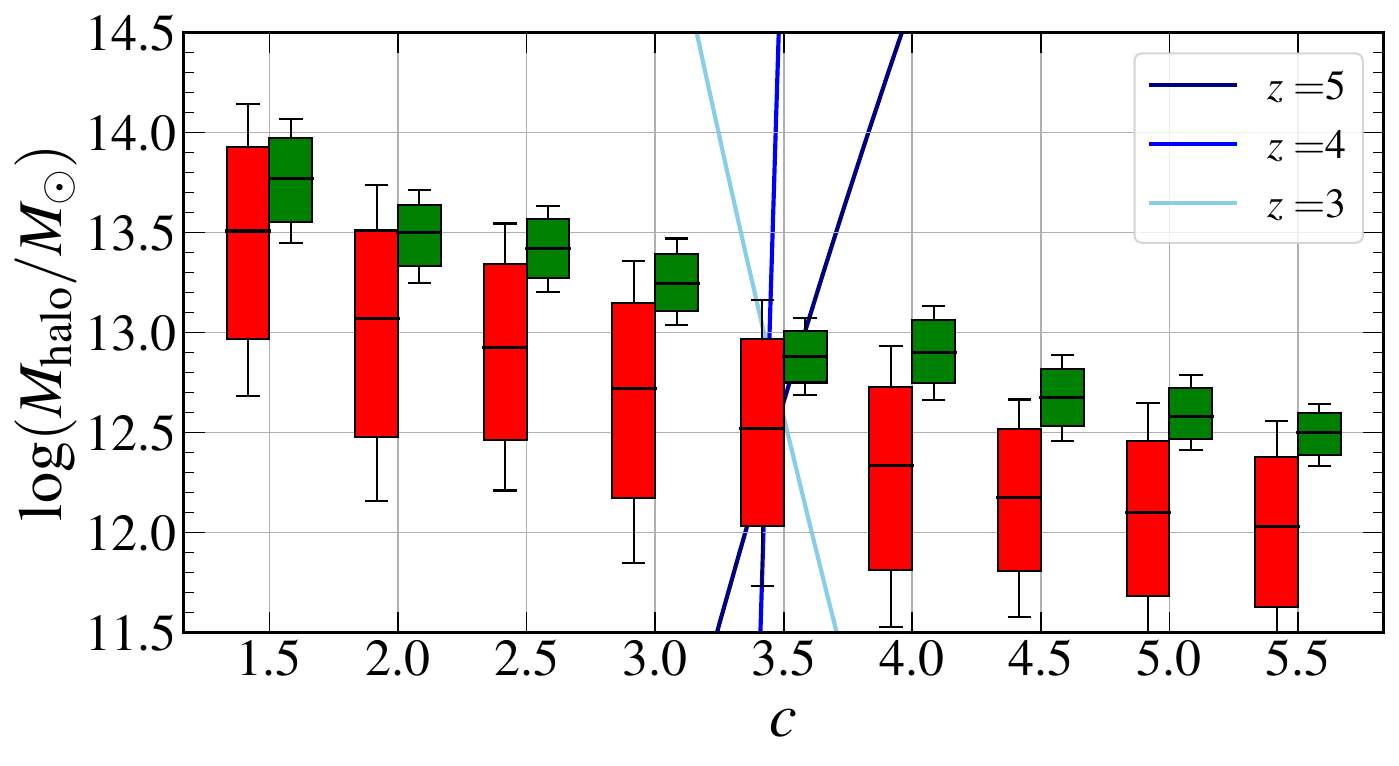}
    \includegraphics[width=0.45\linewidth]{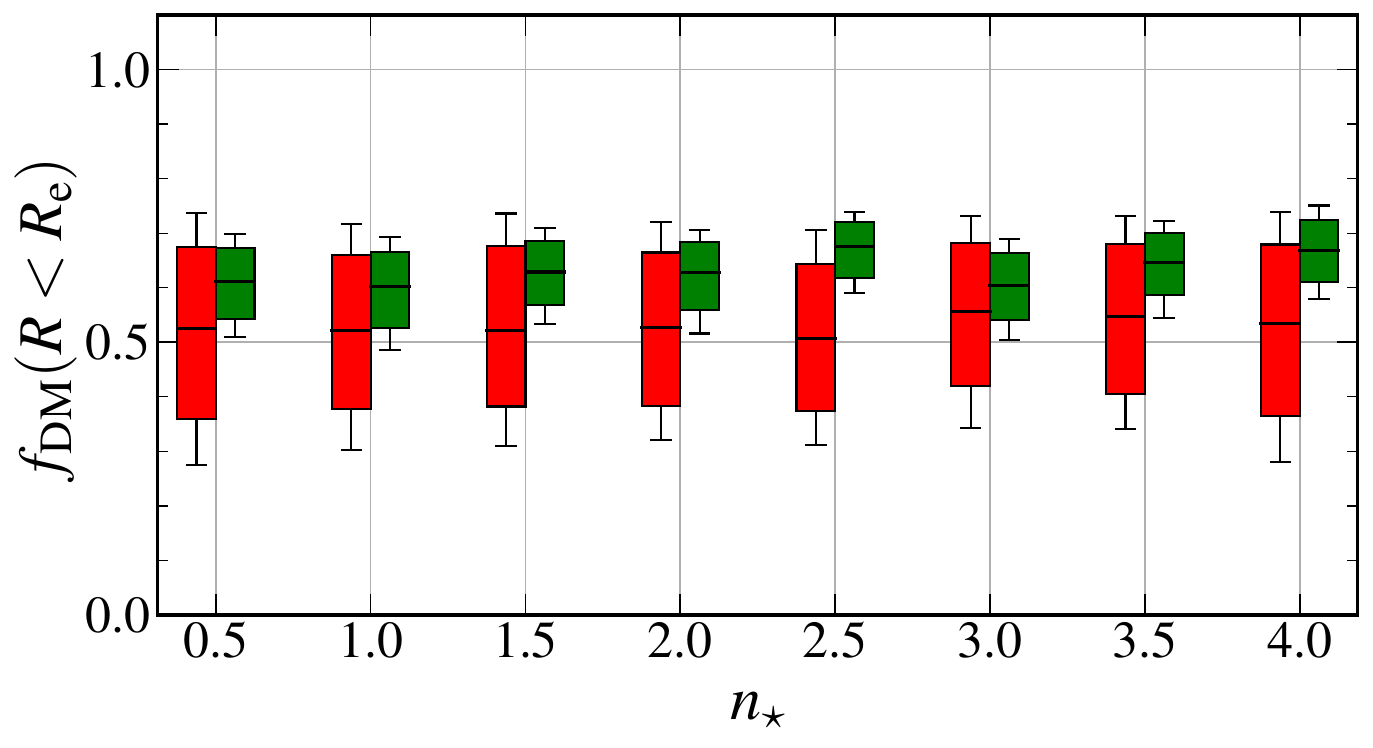}
    \includegraphics[width=0.45\linewidth]{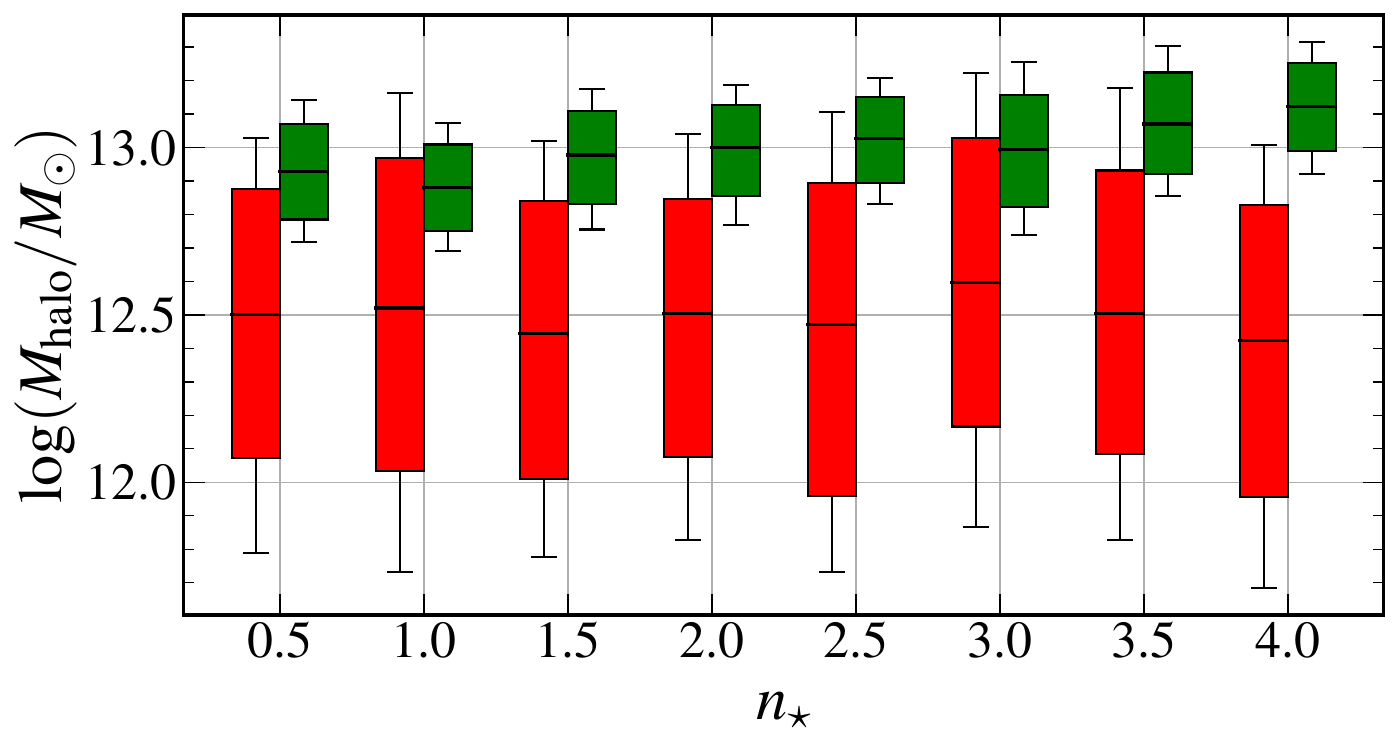}
    \includegraphics[width=0.45\linewidth]{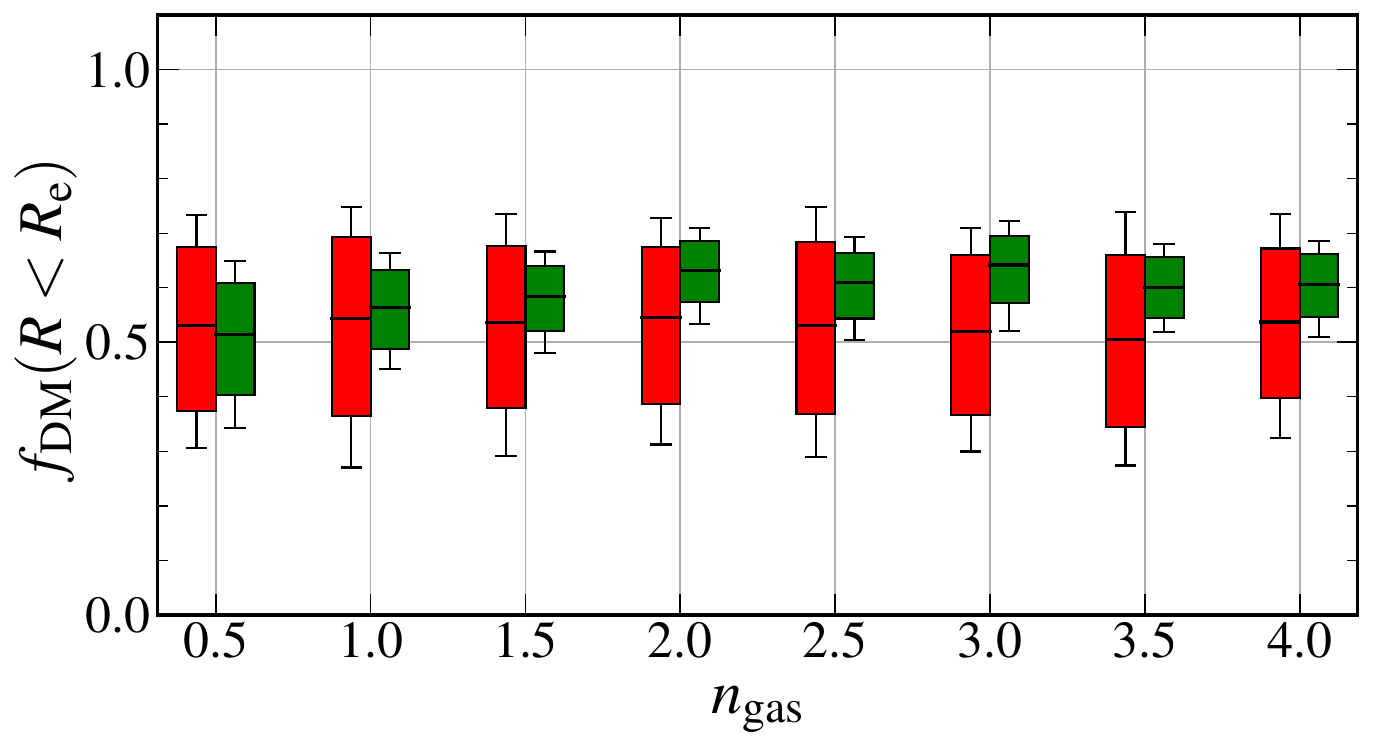}
    \includegraphics[width=0.45\linewidth]{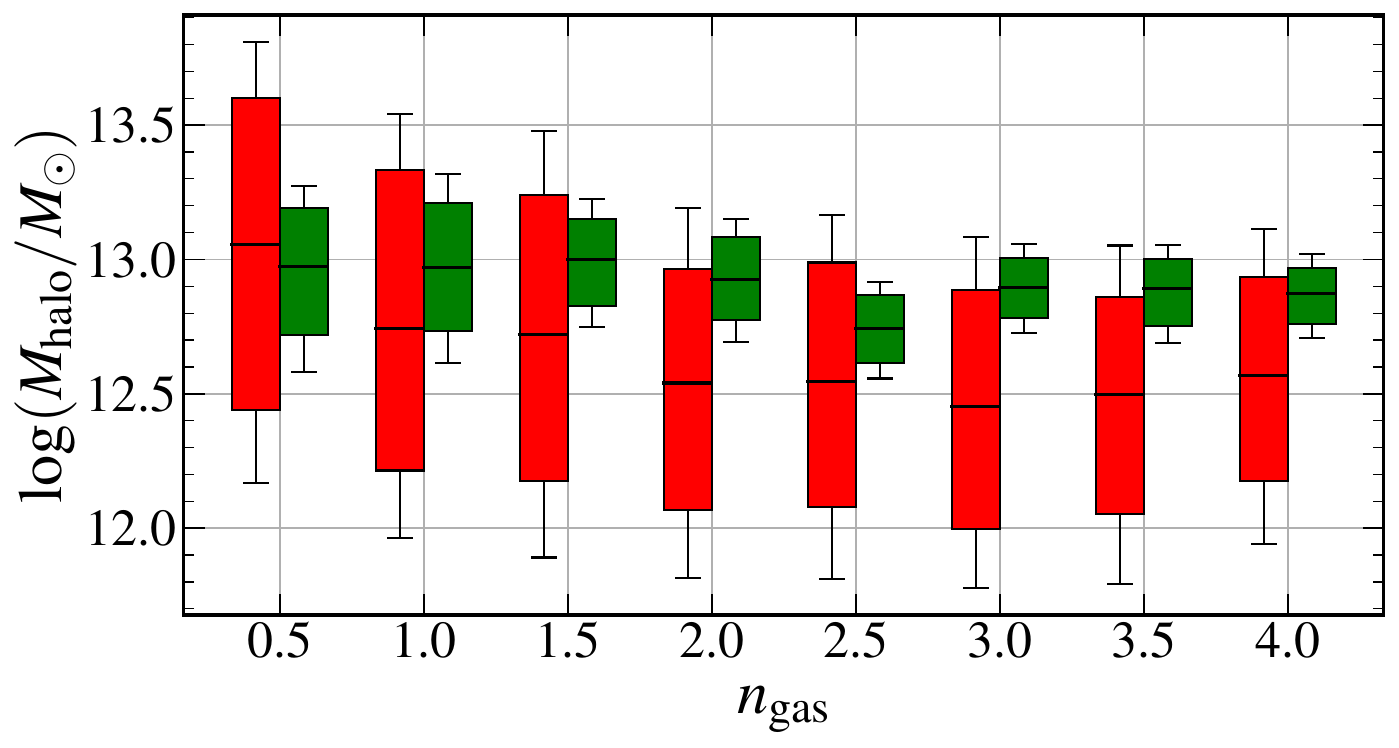}
    \includegraphics[width=0.45\linewidth]{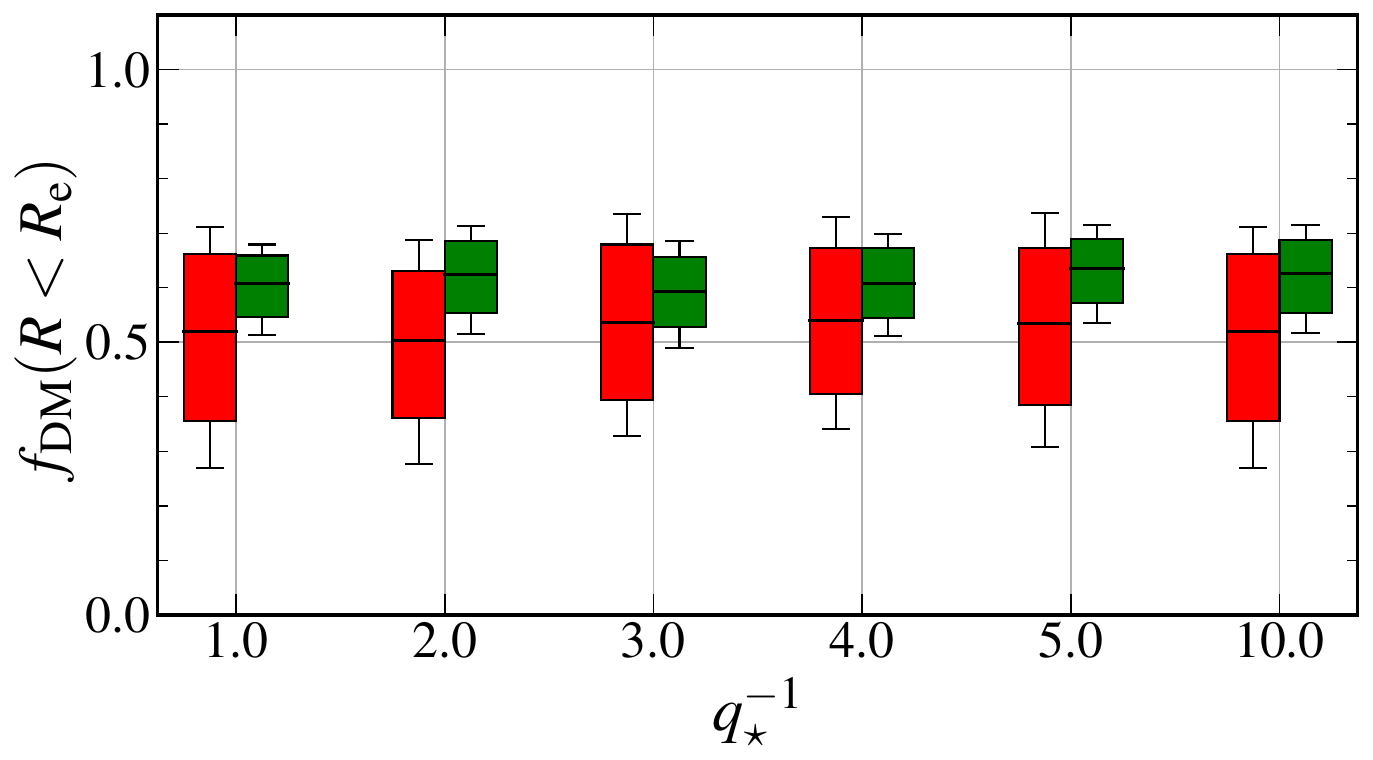}
    \includegraphics[width=0.45\linewidth]{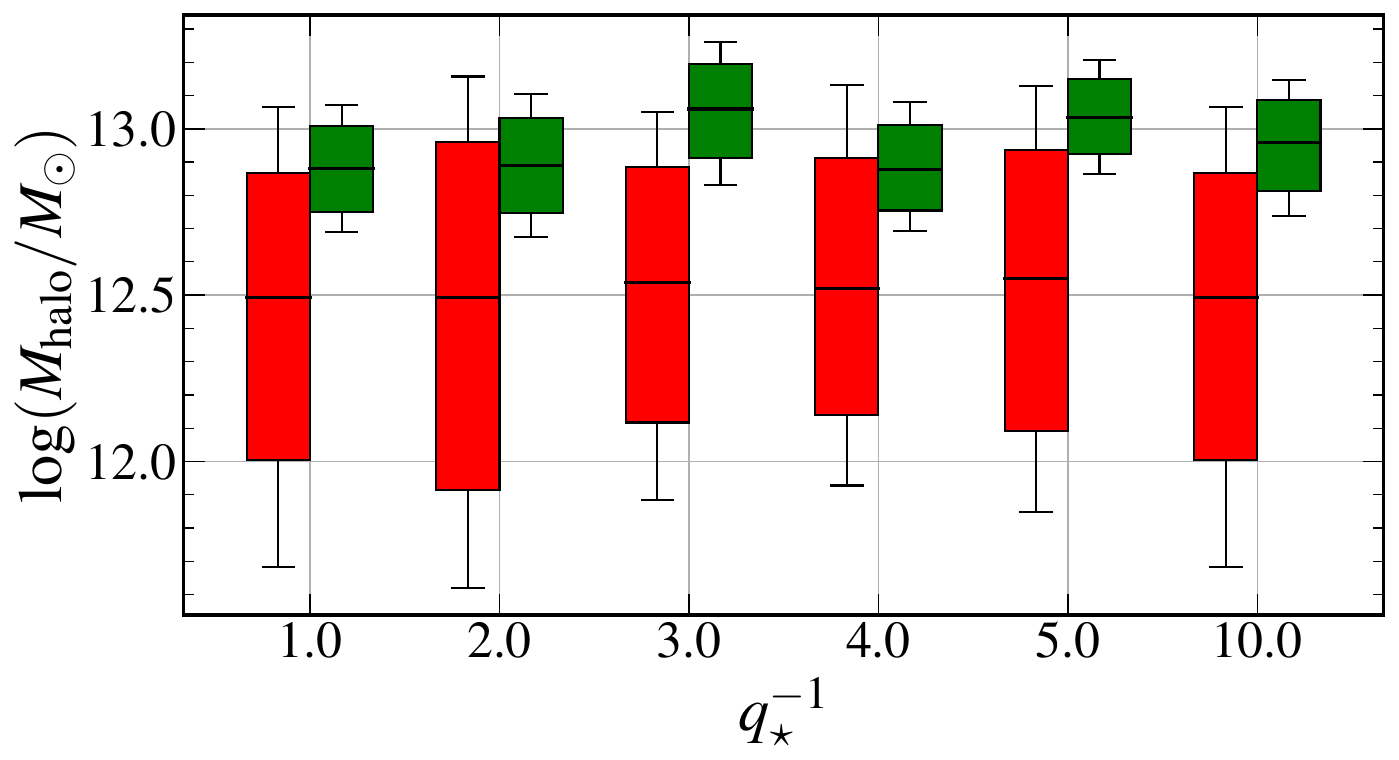}
    \includegraphics[width=0.45\linewidth]{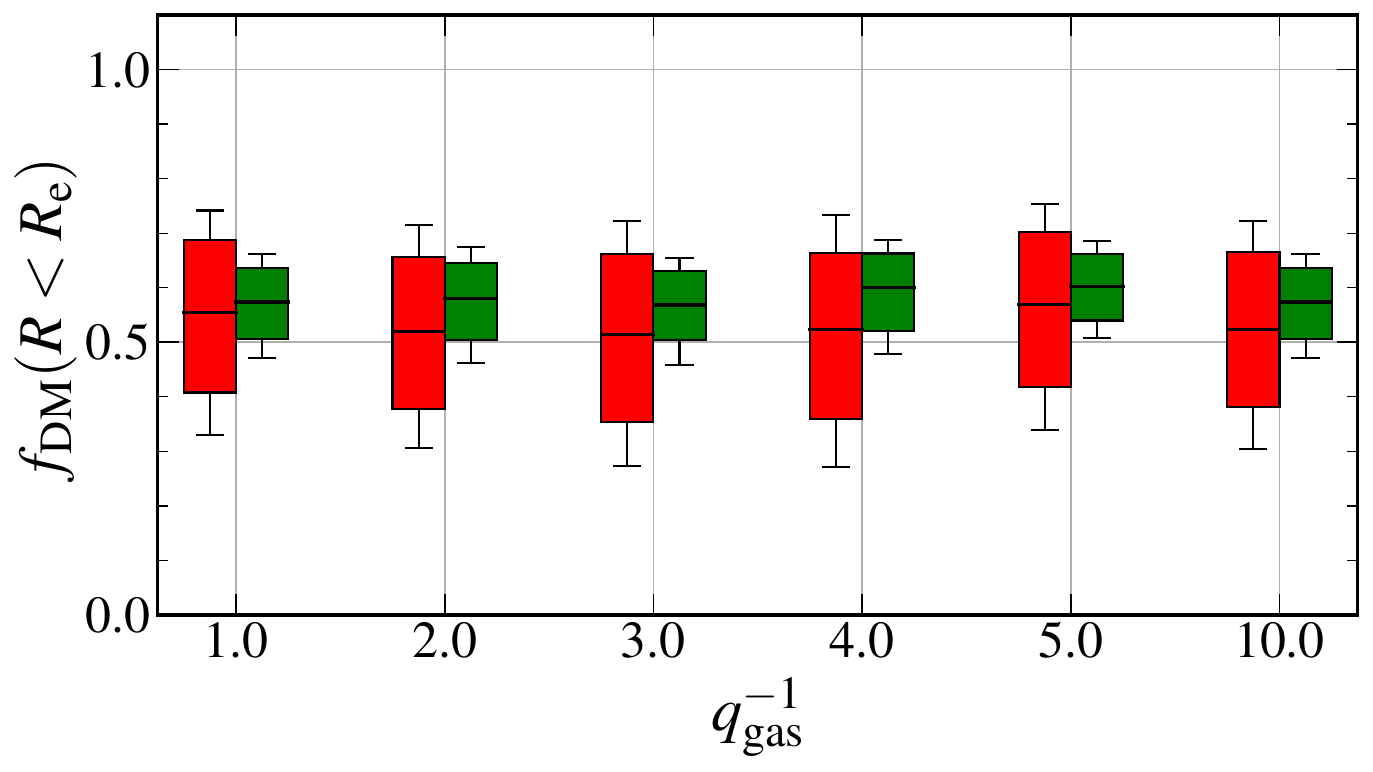}
    \includegraphics[width=0.45\linewidth]{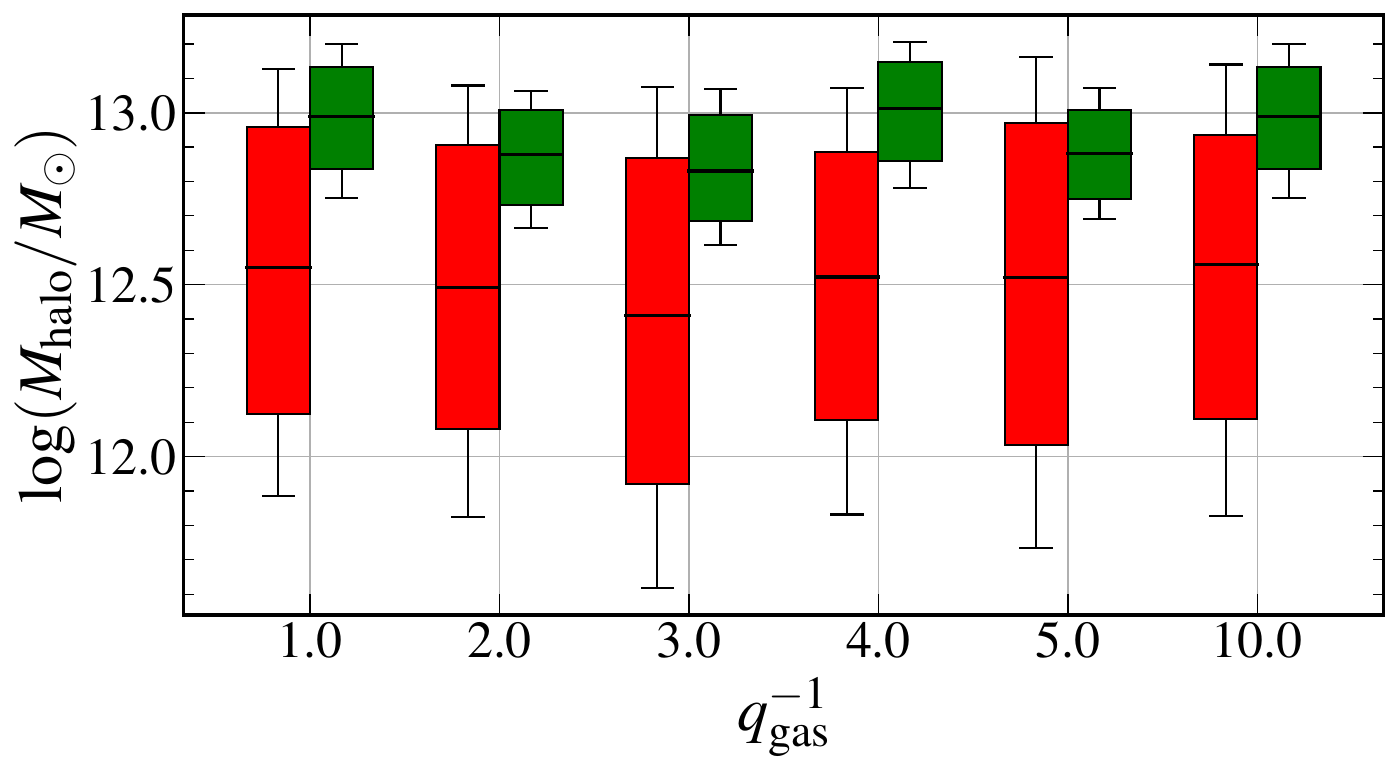}
    \caption{The best-fit dark-matter fraction within the effective radius ({\it left column}) and the virial mass of the dark matter halo ({\it right column}) are estimated by fixing the model parameters to different values. The rows from top to bottom illustrate the effects of varying the halo concentration, the S\'ersic index of the stellar component, the S\'ersic index of the gas component, the thickness of the stellar component, and the thickness of the gas component, respectively.}
    \label{Fig8: para}
\end{figure*}

\cite{Price+2021} point out the importance of the choice of priors when using \dysmalpy, indicating that the fitting results might be sensitive to the selected parameters. This implies that the parameters, especially the S\'ersic index of the stellar component ($n_\star$), utilized in the fitting may be inappropriate since we have not directly observed their light. Previous studies suggest that the stellar component of quasars at this redshift could be an ultra-compact core \citep{Walter+2022} through kinematic analysis, or an extended disk through the {\it JWST} observation \citep{Ding+2023}, suggesting the diversity of stellar distribution of high-$z$ quasars. 
Another important parameter, the halo concentration ($c$), can also contribute to the fitting results. Our current knowledge about the halo concentration all comes from numerical simulations \citep{Dutton+2014, Diemer+2015}, indicating that the concentration is a function of both redshift and halo mass. 
The 3-D geometry of the baryon component, i.e., whether the baryons are distributed in a thin disk or a spheroid, can also influence our fitting result due to the different shapes of potential. Therefore we use the oblate spheroid to model the 3D distribution of both stellar and gaseous components. The thickness of the oblate spheroid is described by the inverse axis ratio ($q^{-1}$), where $q^{-1}=1$ indicates a spheroid and $q^{-1}=10$ represents a flattened disk. 

In order to study how fitting results change with different inputs of parameters ($c, n_\star, q_\star^{-1},$ and $q_{\rm gas}^{-1}$), we conducted \dysmalpy\ fitting with these parameters fixed to various values and examined the resulting parameters of interest. In particular, we examined the dark matter fraction and corresponding halo mass, with $1.5\leq c \leq 5.5$, $0.5\leq n_\star \leq 4.0$, $0.5\leq n_{\rm gas} \leq 4.0$, $1\leq q_\star^{-1}\leq10$, and $1\leq q_{\rm gas}^{-1} \leq 10$. The range of those parameters covers all of the possible parameter space at these redshifts and therefore provides a robust examination of the relationship between parameters of interest and the input parameters. 

The examinations are presented in Figure \ref{Fig8: para}, which illustrates the best-fit \fdm\ and the corresponding $\log M_{\rm h}$ relative to various input parameters. It is shown that the dark matter fraction remains almost constant when different values of $c$, $n_\star$, $q_\star^{-1}$ and $q_{\rm gas}^{-1}$ are chosen, indicating that the dark matter fraction is not affected significantly by these parameters.

The value of $\log M_{\rm h}$ is not a free parameter, but it is linked to \fdm. In Figure \ref{Fig8: para}, it is shown that $\log M_{\rm h}$ decreases significantly as $c$ increases, covering a range of about 2\,dex. This variation may indicate that there are large uncertainties in the halo mass measurements for our targets. In particular, the virial mass of dark matter halo will reach an extremely large value when adopting a small concentration. The reason for this result is that the dark matter mass within the effective radius does not change when adopting different concentrations, while the low concentration suggests hugh fraction of dark matter mass is distributed at a large scale, therefore introducing an unexpectedly large halo mass. When we compare the empirical relationship between $c$, $\log M_{\rm h}$, and redshift, we find that the most likely concentration of the dark matter halo at this redshift is around $\sim 3.5$ \citep{Dutton+2014, Diemer+2015}, leading to a halo mass of $\gtrsim 10^{12.5}\,M_\odot$. We also investigated the relationship between $\log M_{\rm h}$ and key morphological parameters, including the S\'ersic indices of the stellar ($n_\star$) and gas components ($n_{\rm gas}$) as well as the inverse axis ratios of the stellar ($q_\star^{-1}$) and gas components ($q_{\rm gas}^{-1}$) (Fig.~\ref{Fig8: para}). Our analysis reveals that the dark matter halo mass of \targb\ increases by approximately 0.4 dex, with increasing $n_\star$, while the halo mass of \targa\ increases by approximately 0.5 dex with decreasing $n_{\rm gas}$. The halo masses of both targets exhibit minimal sensitivity to variations in the inverse axis ratios of the stellar and gas components. These findings underscore the importance of accurately characterizing the stellar distribution, highlighting the value of \jwst\ observations for this purpose.

We further examined the fitting results by testing various inclination angles. Upon varying the input inclination, the velocity dispersion and \fdm\ remained unchanged. However, the baryon mass and halo mass were influenced by the inclination angle. Specifically, we focused on the fitting results for \targb\ because the inclination angle used in our analysis ($i\approx 40^\circ$) differs from the value reported in the literature \citep[$i\approx 60^\circ$;][]{Neeleman+2021}. To address this issue, we reanalyzed the gas kinematics using \dysmalpy\ across a series of inclination angles, ranging from 30 to 60 degrees in 5-degree increments. This reanalysis yielded a baryon mass in the range of $10^{10.76}-10^{11.10}\,M_\odot$ and a halo mass between $10^{12.6}-10^{13.3}\,M_\odot$. By accounting for the uncertainties associated with the inclination angle, we determined that these uncertainties introduce an additional error of 0.2 dex in the baryon mass and 0.35 dex in the dark matter halo mass. 

\subsection{Comparing with clustering studies}
\label{sec5.2: halo mass}

\begin{figure}
    \centering
    \includegraphics[width=\linewidth]{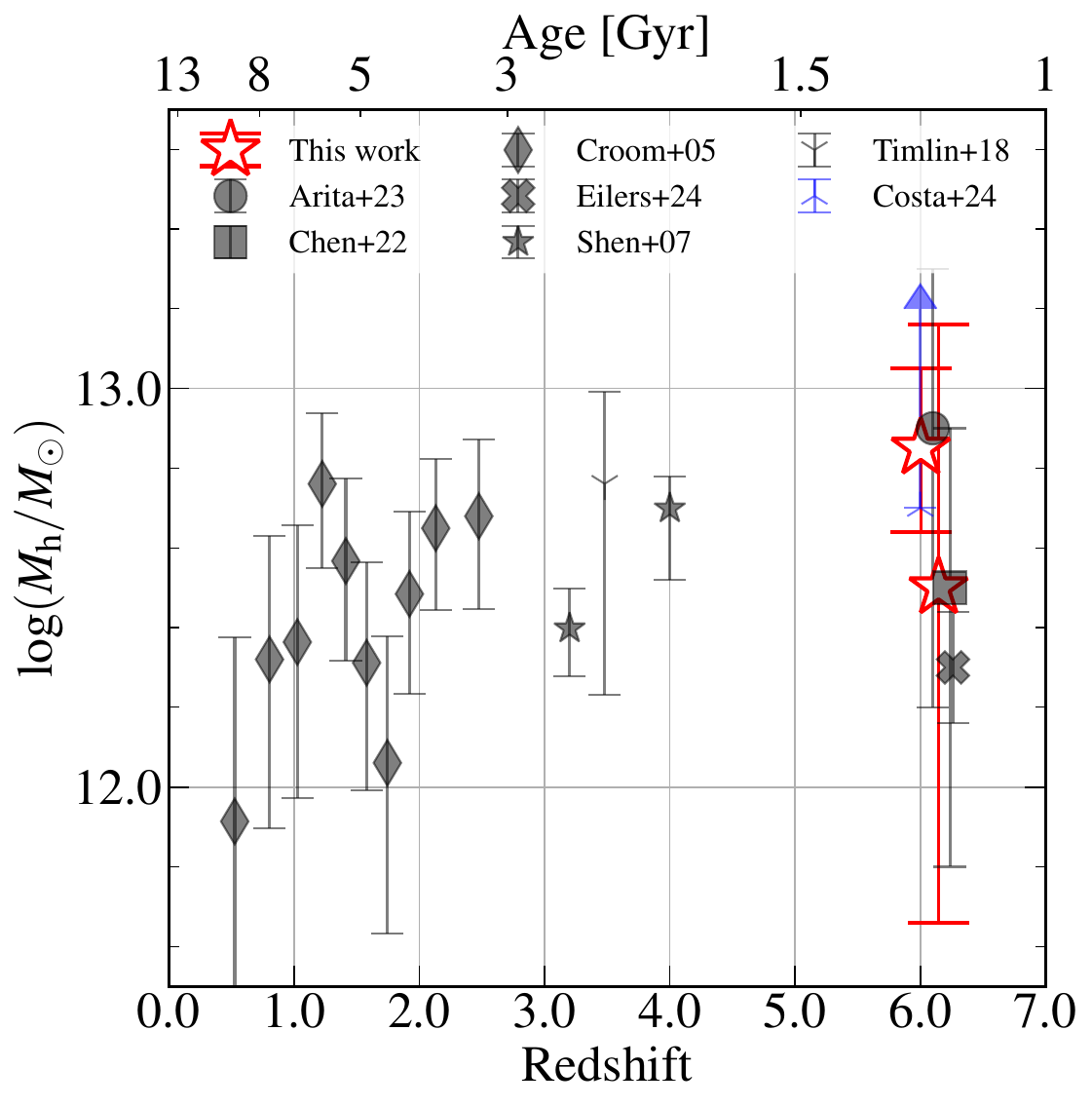}
    \caption{Comparison between the dark matter halo mass of quasars with redshifts. The gray data points represent the halo mass measured through analyzing the galaxy clustering  \citep{Croom+2005, Shen+2007, Chen+2022, Arita+2023, Eilers+2024}, and the blue arrow indicates the lower limit of halo mass as revealed by \cite{Costa2024}. Our measurements are indicated by the red stars.}
    \label{Fig10: halo-mass}
\end{figure}

The large \fdm\ as well as heavy baryonic mass naturally result in a massive dark matter halo (Table \ref{Tab3: dysmal}), which is expected from simulation works \citep{DiMatteo+2005, Springel+2005}. The dense environment and companions provide a large amount of material to feed the growth of the galaxies and SMBHs \citep{Li+2007, DiMatteo+2012}, and eventually make them evolve into the most massive systems in the local universe. Many efforts have been conducted to test this scenario, in particular, to measure the halo mass of those massive SMBH by studying their environments \citep{Arita+2023, Kashino+2023, Wang+2023, Eilers+2024}. 

The halo mass of quasars has been primarily determined by studying the clustering properties of galaxies around them \citep{Arita+2023, Eilers+2024}. This type of analysis has been ongoing for several decades, focusing mainly on quasars at $z\lesssim 4$ \citep{Croom+2005, Shen+2007, Timlin+2018}. In recent years, similar analysis has been extended to quasars at $z\sim 6$, aided by the high-resolution images from the JWST that allow for the separation of companion galaxies and central quasars \citep{Wang+2023, Kashino+2023}. The fundamental approach involves utilizing N-body cosmological simulations to accurately represent the halo clustering properties. Accurate descriptions of the galaxy distribution, e.g., the cross- and auto-correlation function, were conducted based on the simulation or analytic works and then were compared to the observations. It has been observed that the halo mass could vary depending on the simulations or semi-analytical models used  \citep{Behroozi+2019, Schaye+2023, Zhang+2023a, Zhang+2023b}. However, all studies concur that the halo mass of the most luminous quasars is at least $10^{12.3}\,M_\odot$ \citep{Eilers+2024}. There was a study focused on a different quasar sample with relatively lower luminosities \citep{Matsuoka+2018, Matsuoka+2019, Onoue+2019, Chen+2022}. By using the auto-correlation function of approximately one hundred faint quasars at $z \sim 6$, \cite{Arita+2023} inferred a host dark matter halo mass using the linear halo model software \citep{Murray+2013, Murray+2021} and obtain a significantly larger host dark matter halo mass estimate of $10^{12.9}\,M_\odot$. Both studies agreed that the mass of dark matter halos around $z\sim 6$ quasars should be at least $10^{12}\,M_\odot$. 

In addition to studying galaxy clustering, \cite{Costa2024} offered a different method for measuring halo mass by examining galaxy kinematics. They observed that companion galaxies and quasars exhibit relative offset velocities, with the satellites accelerating as they approach the quasar host galaxy. This results in a broadening of the line-of-sight velocities of companion galaxies, which decreases as a function of the distance to the quasar host galaxy. Through simulations, it was determined that the increase in velocity dispersion is especially noticeable when the halo mass is greater than $10^{12.7}\,M_\odot$. 

We compared our measurements of halo mass using gas kinematics to those calculated using other methods, and the results are shown in Figure \ref{Fig10: halo-mass}. Our findings reveal that the halo mass obtained from our method aligns with the values derived from galaxy clustering and galaxy kinematics \citep{Arita+2023, Eilers+2024, Costa2024}. We have to emphasize that our halo mass is extrapolated from the dark matter mass within a relatively small region ($r\lesssim 3-6R_e$), assuming a typical NFW profile with a specific concentration. However, we note that our halo mass and the adopted concentration align with the empirical mass-concentration-redshift relation (Sec.~\ref{sec4.4: paras}; \citealt{Dutton+2014, Diemer+2015}), further supporting that our halo mass does not contradict previous studies. 

\subsection{Relation between $M_{\rm halo}$, $M_{\rm BH}$, $M_{\rm host}$}
\label{5.3: relation}
\begin{figure*}
    \centering
    \includegraphics[width=0.49\linewidth]{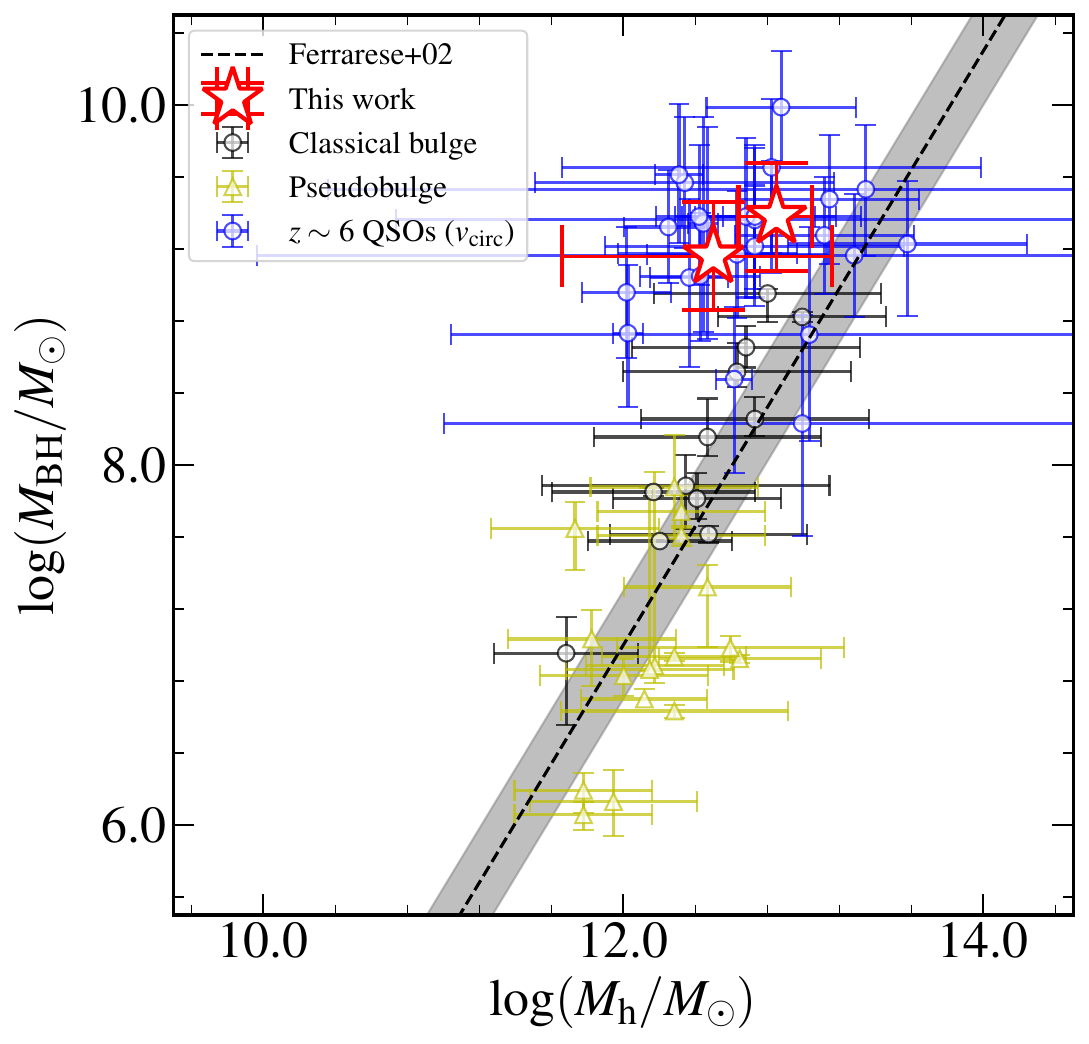}
    \includegraphics[width=0.49\linewidth]{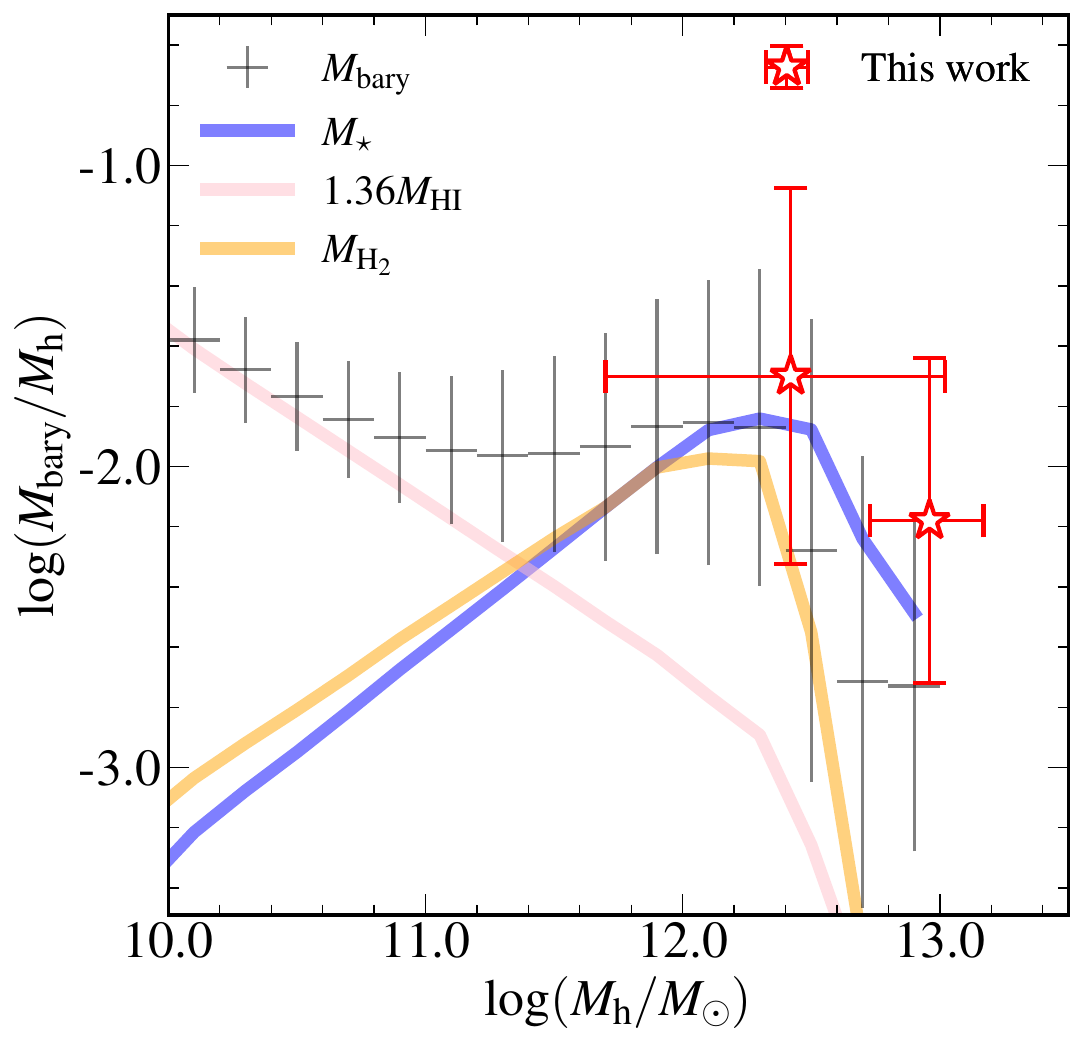}
    \caption{{\it Left panel:} Relation between SMBH mass and halo mass. The black and blue data points represent the local galaxies with classical bulge \citep{Kormendy+2013} and $z\sim 6$ quasars \citep{Neeleman+2021}. The yellow triangles represent the local galaxies with pseudobulge. The red stars denote our measurements, respectively. {\it Right panel:} Baryon mass vs. halo mass for our targets (red stars). The gray errorbars denote the mean values of baryonic-to-halo mass ratio of galaxies at redshift 6 within 0.2\,dex halo mass bins, adopted from the catalog generated from the \texttt{NeutralUniverseMachine} \citep{Guo+2023}. The blue, pink, and orange lines represent the ratio between the stellar mass, the neutral gas mass, the molecular gas mass, and the dark matter halo mass, respectively.}
    \label{Fig11: relationship}
\end{figure*}

The study of local galaxies has revealed a strong correlation between the mass of SMBHs and the mass or velocity dispersion of their host galaxy bulges \citep{Magorrian+1998, Ferrarese+2000, Gebhardt+2000, Kormendy+2013}. These correlations are thought to be established during the active galactic nucleus (AGN) phase, where SMBH activity influences the properties of the interstellar medium, subsequently shaping the overall characteristics of the host galaxy \citep{Silk+1998, DiMatteo+2005, Heckman+2014, Harrison+2018}. However, as research extends to higher redshifts \citep{Bennert+2011, Bennert+2021, Cisternas+2011, Simmons+2011, Schramm+2013, Ding+2020, Ding+2022a, Tanaka+2024}, a notable trend emerges: the SMBH masses at these redshifts are significantly larger than predicted by the relationships observed in the local universe \citep{Ding+2020, Harikane+2023, Maiolino+2023, Ubler+2023}, particularly in the case of the most luminous quasars \citep[e.g., ][]{Shao+2017, Izumi+2019}. Although the BH mass measurements for those high-$z$ quasars might be overestimated \citep{Abuter+2024}, and selection bias may play a significant role in the study of the $M_{\rm BH}-M_\star$ relation \citep[e.g.,][]{Li+2024}, recent deep surveys conducted with the James Webb Space Telescope ({\it JWST}), which offer less biased samples \citep[e.g., ][]{Harikane+2023, Maiolino+2023}, suggest that SMBHs are intrinsically overmassive relative to their host galaxies \citep{Pacucci+2023}.

Beyond stellar mass, \cite{Ferrarese2002} proposed a relationship between SMBH mass and halo mass in local galaxies. However, this tight correlation may arise from the combined effects of the $M_{\rm BH}-M_{\rm bulge}$ and $V_c-\sigma$ relations \citep{Courteau+2007, Ho+2007, Kormendy+2013}. The formation and evolution of SMBHs with masses around $M_{\rm BH} \sim 10^9\,M_\odot$ at redshifts $z\sim 6-7$ remain poorly understood, posing significant challenges to our current understanding of black hole growth and galaxy evolution \citep{Inayoshi+2020}. Though SMBH masses often seem overmassive in luminous quasars compared to dynamical mass \citep[e.g., ][]{Izumi+2019}, they are consistent with local relations for fainter AGNs \citep{Maiolino+2023}. 

\cite{Shimasaku+2019} examined the relationship between SMBH mass and halo mass in $z\sim 6$ quasars, finding only a weak correlation, likely due to significant uncertainties. We revisited this relationship for quasars at $z\sim 6-7$ and for local galaxies, following the methodology used by \cite{Shimasaku+2019}. The black hole masses for the local galaxies were directly taken from \cite{Kormendy+2013}, while those for high-redshift quasars were adopted from \cite{Neeleman+2021}. Specifically, the SMBH mass for \targb\ was estimated from the \civ\ emission line, and that for \targa\ from the \mgii\ emission line. 
We estimated the halo masses for both low-redshift galaxies and high-redshift quasars using the procedure in \cite{Ferrarese2002}. 
The circular velocities ($v_{\rm circ}$) can be used to estimate the virial mass of the dark matter halo by assuming it represents the halo velocity ($v_{\rm vir}$) measured at the virial radius ($R_{\rm vir}$) and the virial mass of the dark matter halo is proportional to the third power of $v_{\rm vir}$ according to the virial theorem. The constant of proportionality depends on the adopted cosmology. In our assumption, we adopted the same $\Lambda$CDM cosmological model used by \citet{Bullock+2001}. 
The circular velocities for the local galaxies were taken from \cite{Kormendy+2013}, and those for high-redshift quasars from \cite{Neeleman+2021}. The halo masses for our two quasars were derived from our kinematic analysis in Sec.~\ref{sec4.4: paras}. The results, presented in the left panel of Figure \ref{Fig11: relationship}, indicate that our targets exhibit lower uncertainties in halo mass. In addition, our targets are close to the local relation between halo masses and SMBH masses. However, we caution that our sample size is extremely limited, with only two targets, making it difficult to draw definitive conclusions about whether black hole masses are regulated by halo mass. This underscores the need for a larger sample encompassing a broader parameter space.

We then evaluated whether the baryonic mass and halo mass derived from our method follows the baryon mass - halo mass relation \citep{Guo+2023}. Using the catalog generated from \texttt{NeutralUniverseMachine},\footnote{\url{https://halos.as.arizona.edu/UniverseMachine/DR1/SFR_ASCII/Gas_Masses_NeutralUniverseMachine/}} we derived the connection between the baryon mass and the halo mass for galaxies at $z\sim 6$. Our targets are situated at the high-mass end of the halo mass function at this redshift \citep{Behroozi+2019}, and are consistent with the empirical model, after taking into account the uncertainties. In the future, a large sample with dynamic measurements of galaxy mass and halo mass is essential for establishing such relationship for massive galaxies at high redshift.

\subsection{Evolution of \fdm}
\label{sec5.1: fdm}

The galaxy mass derived from our kinematic analysis is approximately $10^{10}-10^{11}\,M_\odot$, which is similar to that of massive galaxies at lower redshifts \citep{Genzel+2020, Rizzo+2021}. However, our measurements indicate a relatively larger dark matter fraction within the baryon effective radius (Figure \ref{Fig7: fdm}; \citealt{NestorShachar+2023}), but appears to be similar to that of low-mass galaxies at similar redshifts \citep{deGraaff+2024a,deGraaff+2024b} based on dynamical assessments of their dark matter fraction, suggesting that these galaxies are at an early evolutionary stage. This result is unexpected from the simulation work \citep{deGraaff+2024b}. 

However, as we currently only have two targets, our analysis may be biased by selection and may not represent the entire population of massive galaxies at this redshift. It is important to note that our targets are quasars, which are believed to exist in the densest regions of the universe and consequently result in a higher dark matter fraction. Numerical simulations have suggested that the feedback from quasars may impact the distribution of dark matter halos \citep{Khrykin+2024}.  If our targets are in the earliest stages of evolution, we can anticipate that the strong quasar feedback will influence the final dark matter halo. The redistribution of dark matter from the inner regions to larger radii, along with the rapid growth of the stellar component, may eventually lead our targets to become baryon-dominated systems at low redshift. 

\subsection{\dysmalpy\ fitting with only high-resolution data}
\label{sec5.2: high-res only}
\begin{figure*}
    \centering
    \includegraphics[width=0.49\linewidth]{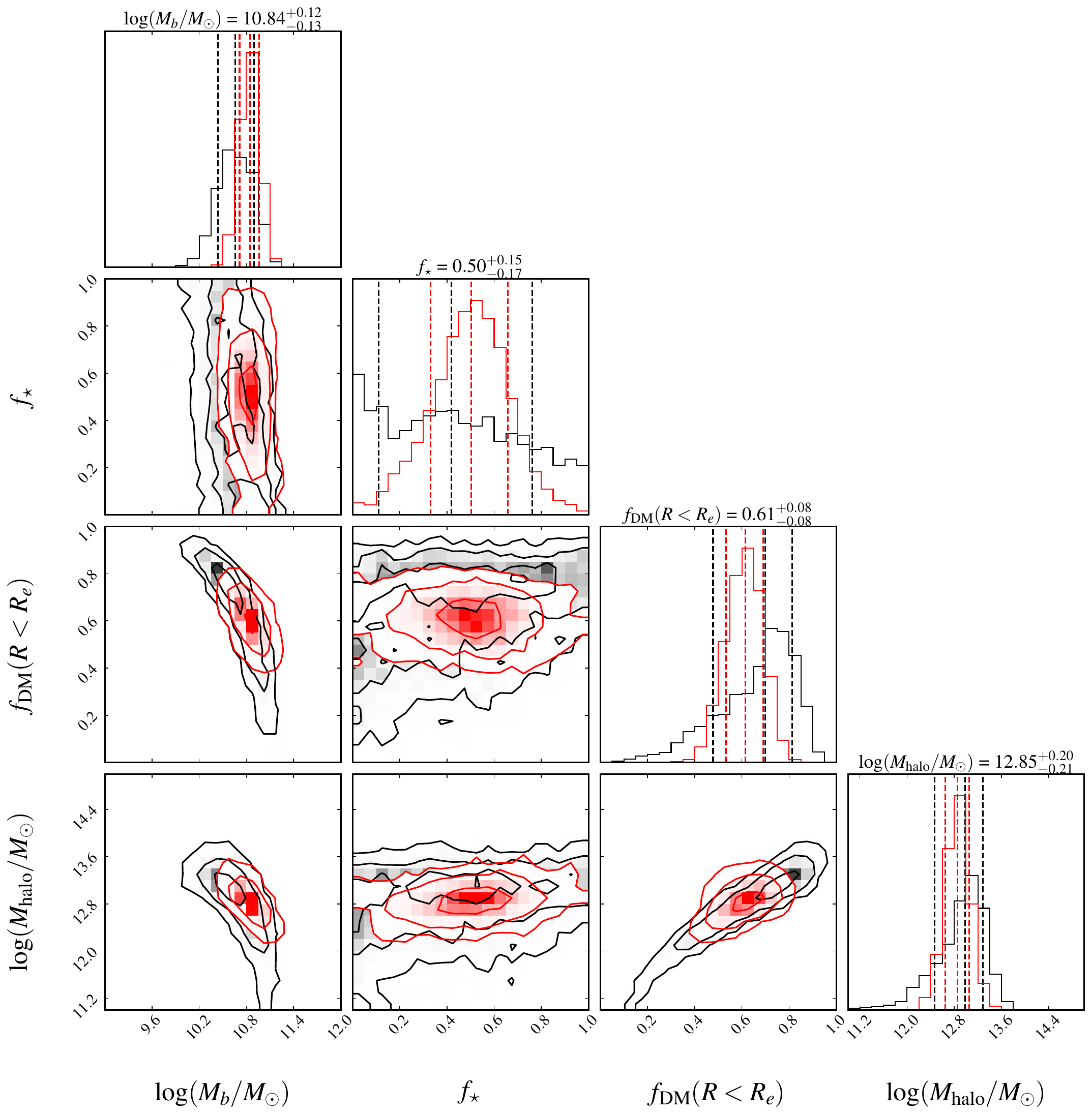}
    \includegraphics[width=0.49\linewidth]{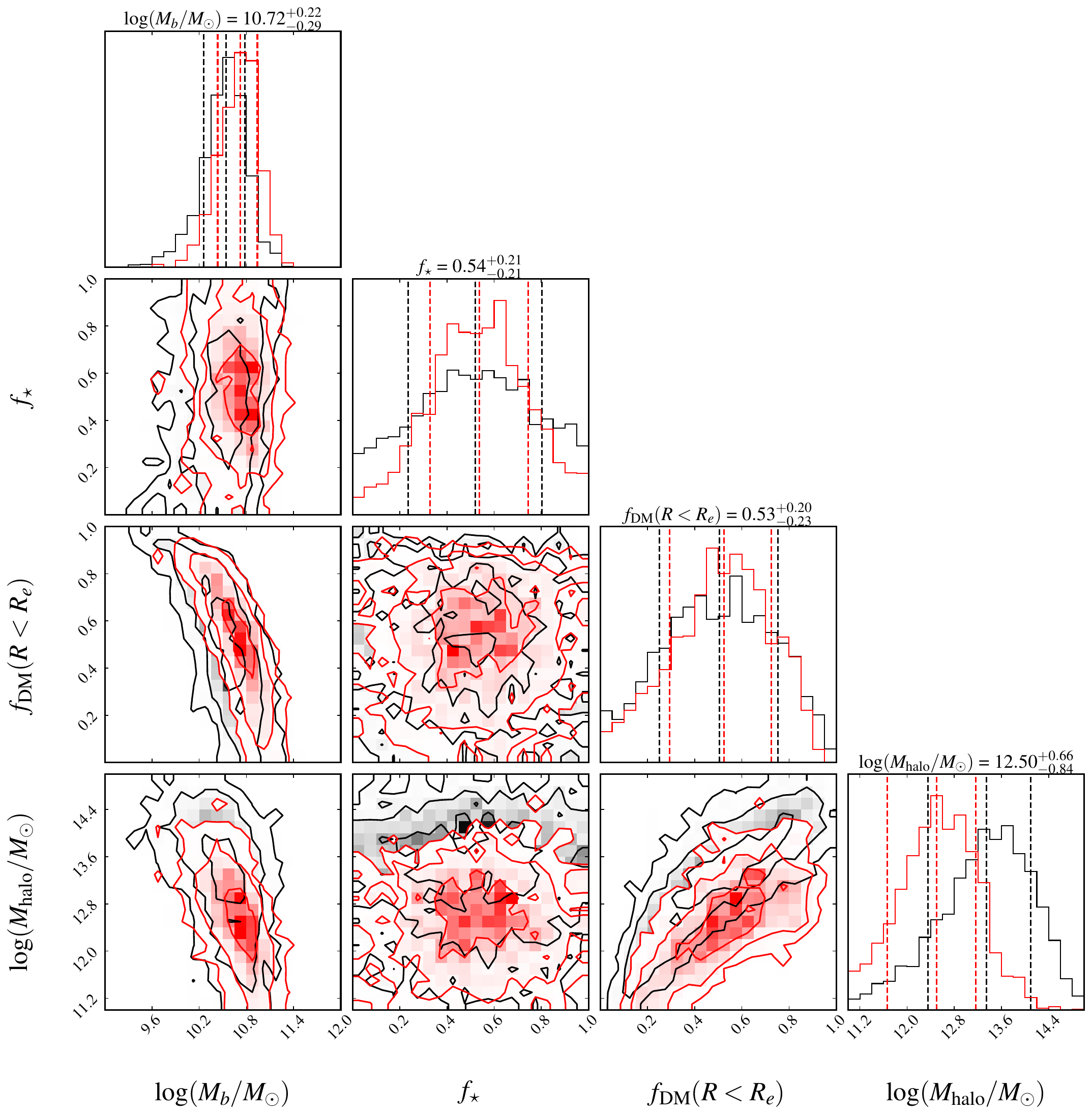}
    \caption{Posterior distribution functions (PDFs) of baryonic mass ($\log M_b$), the ratio between stellar mass and baryon mass ($f_s$), the dark matter fraction within the effective radius \fdm\, and the dark matter halo mass ($\log M_h$) for \targb\ ({\it left panel}) and \targa\ ({\it right panel}). The red contours represent the PDF reported in Sec.~\ref{sec4.2: kinematics}, while the black contours represent the PDF derived from fitting with only high-resolution data.}
    \label{Fig9: corner}
\end{figure*}

\cite{Price+2021} have emphasized the significance of prior constraints and data quality in obtaining accurate parameters. In this section, we examined the effects of using low-resolution data. We conducted the \dysmalpy~ fitting using the method outlined in Sec \ref{sec3.1: dysmalpy}, but this time we only used high-resolution data. The input parameters and mask were set to be the same as in Sec.~\ref{sec3.1: dysmalpy}.

The distributions of $\log M_b$, \fdm, and $\log M_h$ are shown in Figure \ref{Fig9: corner}, the black contours represent the posterior distributions of parameters using only high-resolution data, and the red contours represent that of parameters using both high- and low-resolution data. The obvious anticorrelation between $\log M_b$ and \fdm\ illustrates the degeneracy between these two parameters \citep{Price+2021}. 

For \targb, we note that although the peak of the red histograms is close to that of the black histograms, with a slight offset, suggesting that the values of the best-fit parameters derived from two observational datasets are consistent with that derived from the high-resolution data, considering the uncertainties. However, we can obviously notice that distributions obtained from the full datasets span a narrower range compared to that estimated with only high-resolution data, and have smaller uncertainties. For \targa, the PDF of $\log M_b$ does not change significantly, while we find that the distribution of \fdm\ is significantly improved with the two datasets. In particular, the distribution of \fdm\ derived from only high-resolution data almost spans all of the parameter space, with no obvious peak. After adding the low-resolution observation, the distribution of \fdm\ of this target peaks at around 0.5, and strongly disfavors the parameter space of $\lesssim 0.2$, suggesting that this target is a dark matter dominated system. This result significantly reduces the probability of $\log M_h\lesssim 10^{12}\,M_\odot$, implying that this target is unlikely located within a light halo ($\log M_h<10^{12}\,M_\odot$), which cannot be ruled out with only high-resolution data.

\cite{Price+2021} pointed out that the Gaussian priors on $\log M_b$ will help break the \fdm--$\log M_b$ degeneracy by restricting the posteriors of these parameters to narrower ranges, suggesting the importance of estimating the galaxy mass from other methods, e.g., from spectral energy distribution (SED) \citep{Conroy+2013}, which can now be performed for high-$z$ galaxies, or even low-luminous quasars (e.g., \citealt{Ding+2023, Fujimoto+2024}). However, estimating the stellar mass of luminous quasars using SED at $z\sim 6$ is still difficult since its emission is dominated by the central AGN and the stellar light is too faint to be detected \citep{Stone+2024, Yue+2024}, which means it is quite hard to break the degeneracy between $\log M_b$ and \fdm\ when applying \dysmalpy\ on luminous quasars. However, we noticed that the new low-resolution data can also help to narrow the parameter space and provide a more accurate measurement. The low-resolution data, which traces the gas kinematics at the outskirts of galaxies where the gravitational potential is dominated by the dark matter halo, makes it much easier to infer the properties of dark matter compared to using only high-resolution data. The low-resolution data effectively reduces the degeneracy between the baryon and dark matter mass, highlighting the significance of such observations. 

\subsection{Final considerations}
\label{sec5.4: possibilities}

All of the above analyses and discussions are based on the assumption that the \cii\ emission traces a gravitationally dominated disk. This assumption requires that the gas motion is solely influenced by the gravitational potential, allowing us to estimate the mass distribution based on the kinematics. However, it is crucial to recognize that quasars at $z\sim 6$ may be experiencing various activities, such as recent mergers \citep{Banados+2019, Decarli+2019, Decarli+2024, Izumi+2021, Izumi+2024}, AGN-driven outflows \citep{Cicone+2015, Carniani+2015, Richings+2018}, and a combination of these two effects \citep{Tripodi+2024}. These processes could significantly influence the velocity gradients observed in our targets, potentially complicating the interpretation of our results. Consequently, we cannot entirely rule out the possibility that the observed kinematics are affected by these effects rather than purely by gravitational forces. To distinguish between these scenarios and improve the accuracy of our mass estimates, deeper ALMA observations with higher spatial resolution or direct observations of stellar light from quasar host galaxies are needed. Such observations would allow us to disentangle the contributions of rotation, outflows, and merger-driven dynamics. 

In particular, \targb\ is notable for exhibiting two \cii\ peaks separated by approximately 1.5\,kpc in high angular resolution observations \citep{Novak+2020, Neeleman+2021}. This feature could indicate an ongoing merger \citep[see Fig. 2 in ][]{Neeleman+2021}, or the presence of gravitationally unstable clumps \citep{Tadaki+2018}. To assess whether the gravitational potential in the merger scenario is primarily dominated by dark matter, we used archival data, applying the same reduction parameters as described by \cite{Novak+2020}. 
For this merger scenario, we assume that the galaxy is supported by random motions and estimate the dynamical mass using the virial theorem, assuming that the distribution of molecular gas is spherical with uniform density and has an isotropic velocity dispersion. Under these assumptions, the dynamical mass is estimated as $M_{\rm dyn}=5\sigma^2R/G$ \citep{Pettini+2001}, where $\sigma={\rm FWHM}/2.35$ is the one-dimensional velocity dispersion, and $G$ is the gravitational constant. By measuring the FWHM of the \cii\ emission spectra from these clumps and estimating the size of the \cii\ emitting region, we found values of $342\pm20\,\rm km\,s^{-1}$ and approximately 0\farcs5, corresponding to roughly 3\,kpc at this redshift. 
Based on the circular velocity, the dynamical mass of this central region is approximately $3.7\times 10^{10}\,M_\odot$. 
This value is comparable to the dynamical mass within the central 3\,kpc according to the current mass model ($M_{\rm dyn}=V_c^2R/G\approx 3\times 10^{10}\,M_\odot$). The consistency between the dynamical mass estimates under the merger hypothesis and our model suggests that the presence of a merger does not significantly alter our overall conclusions about the mass distribution in \targb.

Some previous studies have commonly used \cii\ luminosity as a tracer of molecular gas mass. To explore this approach, we estimated the gas mass within the central 3\,kpc using the measured \cii\ flux of $2.9\pm0.2\,\rm Jy\,km\,s^{-1}$ and applying the \cii-to-gas conversion factor (\acii$=31\,M_\odot\,L_\odot^{-1}$) from \cite{Zanella+2018}. This yields a gas mass of approximately $8.8\times 10^{10}\,M_\odot$, which is roughly twice the estimated dynamical mass. This overestimation of gas content aligns with findings from \cite{Neeleman+2021}, suggesting that the \acii\ factor derived from low-redshift star-forming galaxies may not be appropriate for high-redshift quasars \citep{Decarli+2022}. To address this discrepancy, we also estimated the gas mass using a range of $\alpha_{\textsc{[Cii]}}$ ($5-20\,M_\odot\,L_\odot^{-1}$), which are more appropriate for high-$z$ starburst galaxies \citep{Rizzo+2021}. The resulting gas mass falls between $1.5-6.0\times 10^{10}\,M_\odot$. This analysis underscores the challenges associated with accurately measuring gas mass using \cii\ observations alone, particularly for high-redshift quasars where the physical conditions may differ significantly from those in local galaxies.

\section{Conclusions}
\label{sec6: conclusion}

We conducted a kinematic analysis of the gas motions in two quasars above redshift of 6, using both high- and low-resolution ALMA observations of \cii\ emission. This allowed us to assess the kinematics at both the scale of the host galaxy and the surrounding gas. By employing two kinematic modeling tools, \dysmalpy\ and \bbarolo, we measured the rotation velocities and velocity dispersion of the molecular gas within these quasar host galaxies. Subsequently, we determined the mass distribution of our targets by decomposing the RCs. Our key findings are as follows:

\begin{itemize}

    \item Our quasar hosts exhibit disk rotation with $V/\sigma\sim 2$ and flattened rotation curves at their outskirts. 
    In one case (\targb), a rising rotation curve out to 8 kpc well beyond the likely stellar emission after filtering out a distinct spatial and kinematic component.

    \item From dynamical modeling and decomposition of the rotation curves, we find that our $z\sim 6$ quasars are dark matter-dominated systems, with the dark matter fraction (\fdm = $0.61_{-0.08}^{+0.08}$ and $0.53_{-0.23}^{+0.21}$, respectively) exceeding what was expected based on studies of lower redshifts. Our assessment of \fdm\ remains consistent and unaffected by variations in the parameters we used for evaluation.
   
    \item The relatively larger \fdm\ may be due to the fact that the quasars at $z\sim 6$ are located in the most massive dark matter halos at that redshift. Our dynamical measurement indicates that these quasars are hosted by a dark matter halo with $\log (M_h/M_\odot) = {12.85_{-0.21}^{+0.20}}$ and $12.50_{-0.84}^{+0.66}$, respectively, which is expected to be the most massive halo and introduces a galaxy overdensity of $5\sigma$ based on simulation work and is consistent with previous studies.
    
    \item Comparing the SMBH mass with the dark matter halo mass determined from our dynamical fitting, we show that quasars might follow the local $M_{\rm BH}-M_{\rm h}$ relation while the SMBHs are overmassive compared to the stellar mass of their host galaxies. This may indicate a heightened role for the halo in growing the first SMBHs. 

\end{itemize}

In conclusion, by studying the \cii\ kinematics, we obtained the mass distribution of two quasar host galaxies and separated the baryonic and dark matter components by the decomposition of the RCs. Our results indicate that both of our quasars are dark matter-dominated systems and are located in the most massive halos at that redshift, suggesting that the dark matter halo plays a crucial role in the formation of early galaxies and the growth of the first SMBHs.

\begin{acknowledgments}
We are grateful to the anonymous referee for constructive comments and practical suggestions that greatly improved this work. 
We acknowledge Prof. P. Behroozi for sharing his knowledge about dark matter halos and for valuable discussions. 
We acknowledge Prof. M. Takada for insightful discussions about the dark matter halo properties at high redshift. 
We acknowledge Dr. B. Kalita for discussions about the ALMA data reduction. 
This project has received funding from NASA through the NASA Hubble Fellowship grant HST-HF2-51505.001-A awarded by the Space Telescope Science Institute, which is operated by the Association of Universities for Research in Astronomy, Incorporated, under NASA contract NAS5-26555. 
LCH and RW were supported by the National Key R\&D Program of China (2022YFF0503401), the National Natural Science Foundation of China (11991052, 12233001), and the China Manned Space Project (CMS-CSST-2021-A04, CMS-CSST-2021-A06). 
RW acknowledges support from the National Natural Science Foundation of China (NSFC) with grant No. 12173002. 
SC acknowledges support by European Union’s HE ERC Starting Grant No. 101040227. GCJ acknowledges funding from the ``FirstGalaxies Advanced Grant from the European Research Council (ERC) under the European Union’s Horizon 2020 research and innovation programme (Grant agreement No. 789056).
GCJ and RM acknowledge support by the Science and Technology Facilities Council (STFC) and by the ERC through Advanced Grant 695671 ``QUENCH. 
N.M.F.S. and J.M.E.S. acknowledge funding by the European Union (ERC Advanced Grant GALPHYS, 101055023). Views and opinions expressed are, however, those of the author(s) only and do not necessarily reflect those of the European Union or the European Research Council. Neither the European Union nor the
granting authority can be held responsible for them.
This paper makes use of the ALMA data: ADS/JAO. ALMA\#2015.1.01115.S, \#2017.1.01301.S, \#2018.1.00908.S, \#2021.1.01320.S. ALMA is a partnership of the ESO (representing its member states), NSF (USA) and NINS (Japan), together with NRC (Canada), MOST and ASIAA (Taiwan), and KASI (Republic of Korea), in cooperation with the Republic of Chile. The Joint ALMA Observatory is operated by the ESO, AUI/NRAO, and NAOJ. 
\end{acknowledgments}

\software{Astropy \citep{Astropy+2013}; CASA \citep{McMullin+2007}; \dysmalpy \citep{Price+2021}; \bbarolo \citep{DiTeodoro&Fraternali2015}}

%

\vspace{5mm}




\appendix
\section{JvM Correction}
\label{secA}
\begin{figure*}
    \includegraphics[width=\linewidth]{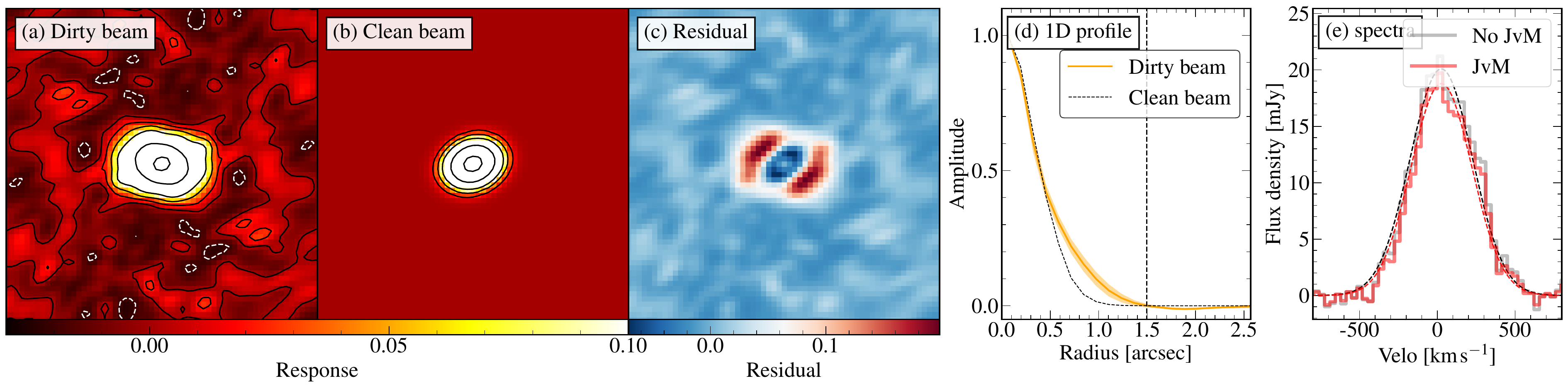}
    \includegraphics[width=\linewidth]{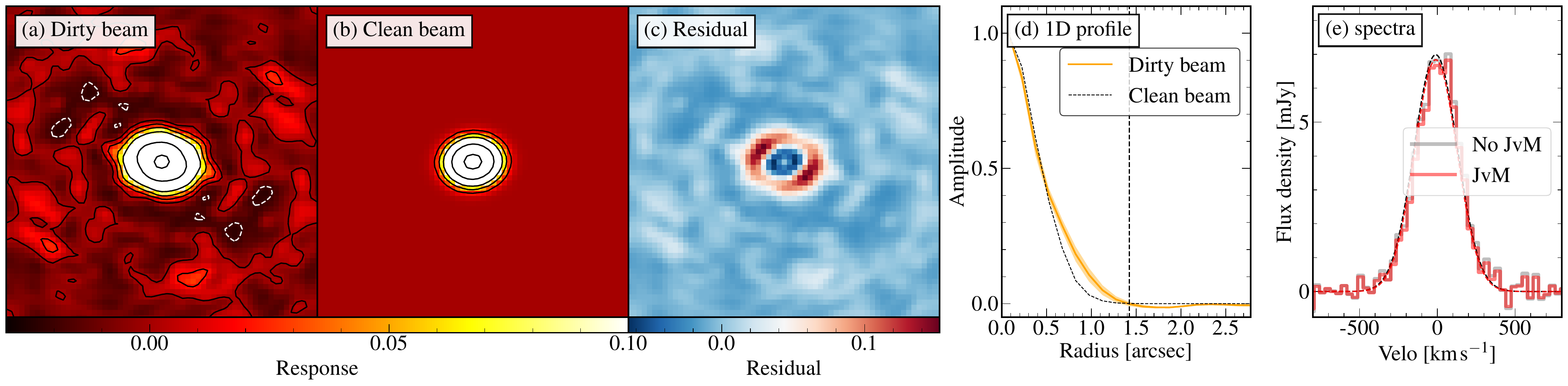}
    \caption{Comparison of the synthesized beam and data products before and after applying the JvM correction. The top row indicates \targb\ and the bottom row illustrates \targa. (a) The sky response (dirty beam) recovered from the CASA \texttt{CLEAN} process, with contour levels at -0.02, 0, 0.02, 0.05, 0.10, 0.50, and 0.80. (b) The clean beam, obtained through CASA  \texttt{CLEAN} task, representing a Gaussian approximation of the main lobe of the dirty beam. (c) The residual between the dirty beam and the clean beam. (d) The azimuthally averaged profiles of both the dirty beam and the clean beam model. (e) The 1D spectra extracted from the JvM-corrected data cube (red histogram) compared to the spectra extracted from the uncorrected data. The red and black dashed lines represent the best-fit 1D Gaussian models for the corrected and uncorrected spectra, respectively.}
    \label{fig13: JvM}
\end{figure*}

\begin{figure*}
    \includegraphics[width=\linewidth]{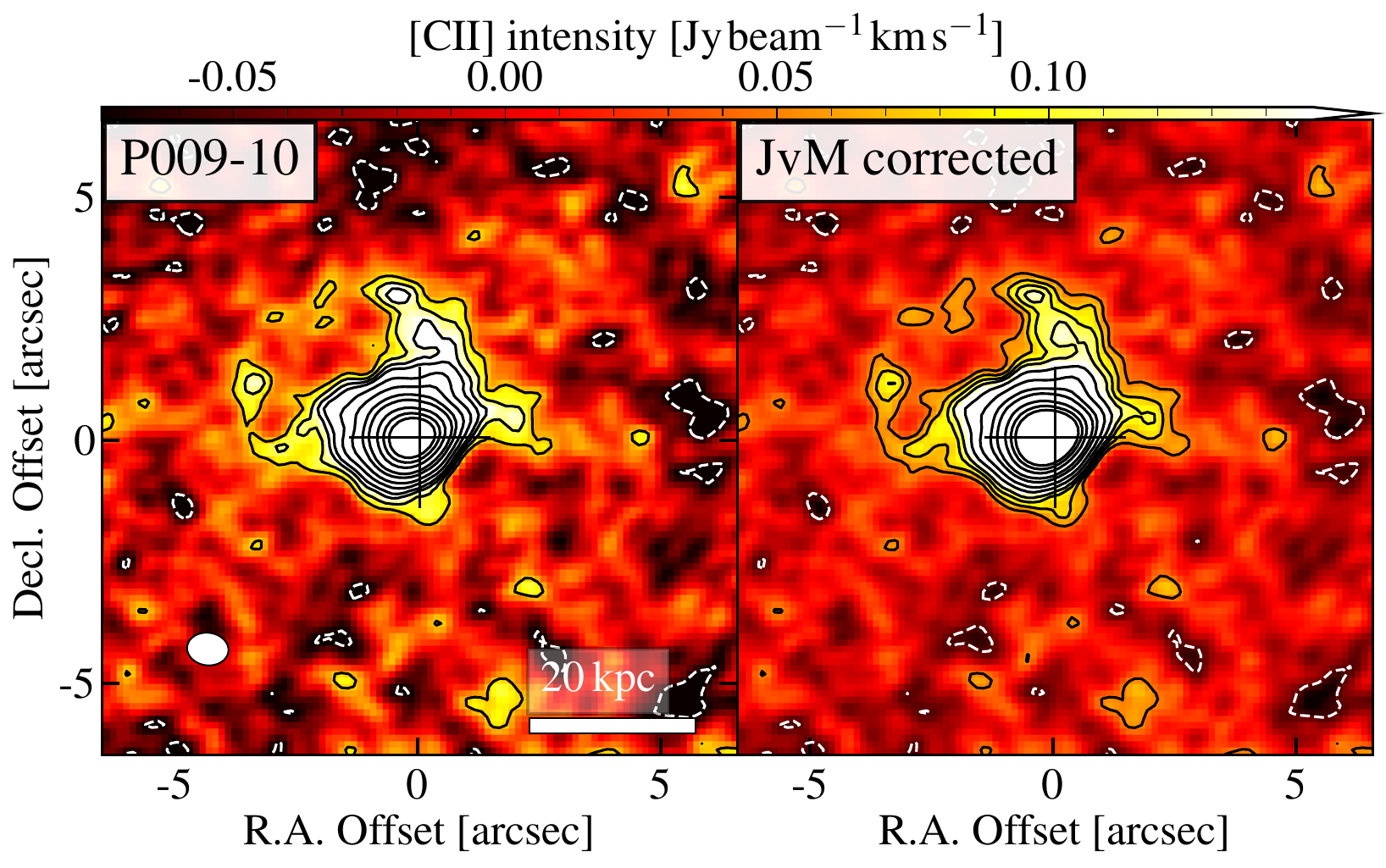}
    \includegraphics[width=\linewidth]{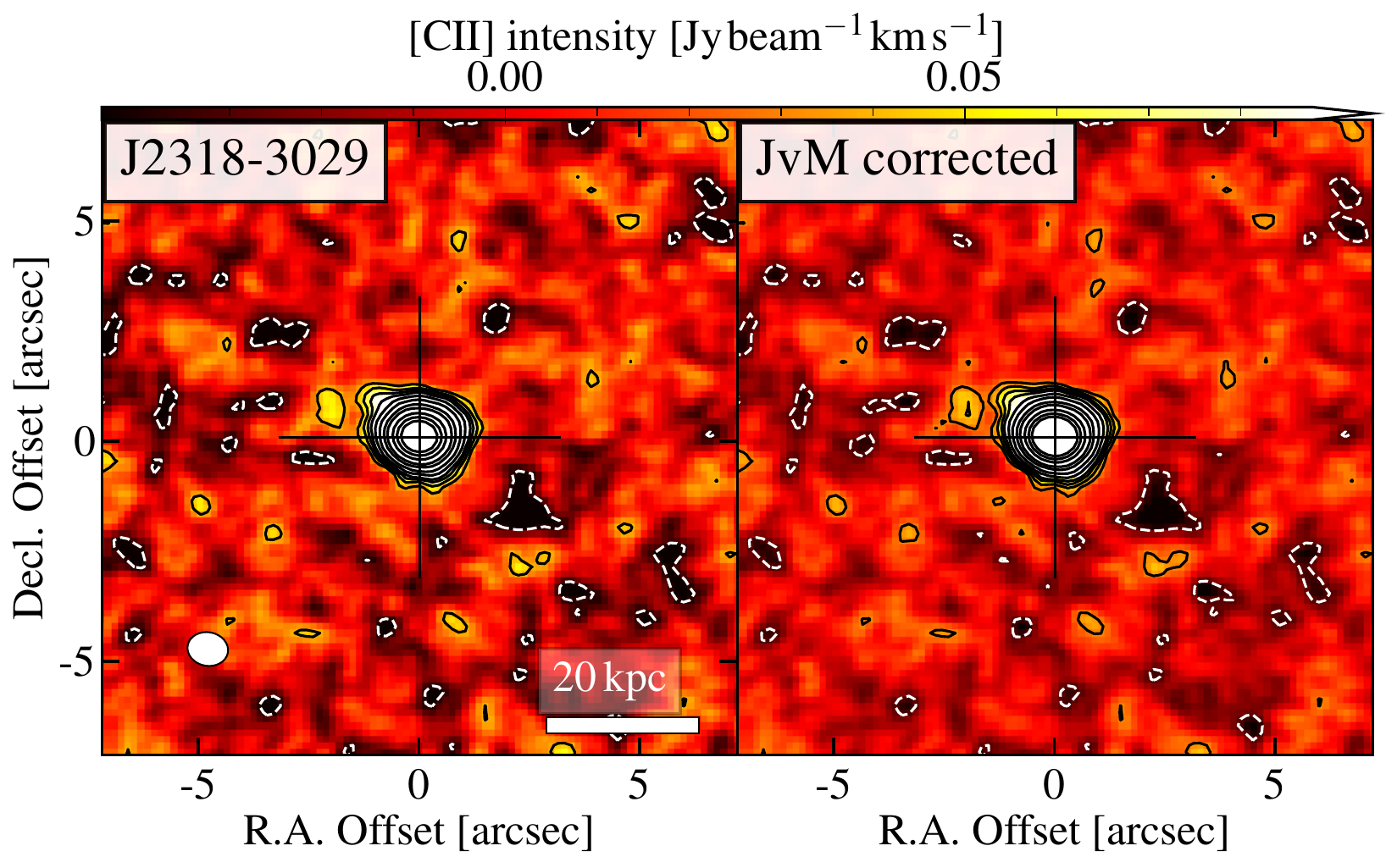}
    \caption{The left column displays the \cii\ intensity maps of our targets produced using the default \texttt{clean} procedure, while the right column shows the \cii\ intensity maps after applying the JvM correction. Contours in each panel represent levels of [-2, 2, 3, 4, 5, 7, 10, 15, 20, 30, 40, 50]$\times \sigma$, where $\sigma$ denotes the root-mean-square (rms) noise in the line-free regions of the intensity map.}
    \label{fig14:comp}
\end{figure*}

As outlined by \citet{Jorsater+95} and \citet{Czekala+21}, the combination of interferometric observations can result in a non-Gaussian synthesized beam due to non-uniform $uv$-coverage, potentially leading to an overestimation of flux measurements. To address this issue and better recover the \cii\ flux and its distribution, we applied the JvM correction following the methodology described by \citet{Czekala+21}. While the detailed algorithm and procedure are thoroughly presented in their work, we provide a brief overview of our implementation of this correction on our data.  

Following this procedure, we constructed a model of the clean beam and compared it to the dirty beam (Fig. \ref{fig13: JvM}). The JvM correction factor was determined by calculating the ratio of the volumes of the clean and dirty beams within their first nulls ($\epsilon = V_{\rm clean beam}/V_{\rm dirty beam}$), yielding values of 0.65 for \targa\ and 0.76 for \targb. This correction factor was then applied to the original residuals. The corrected residuals were added to the convolved model, which represents the clean model convolved with the clean beam. By comparing the 1D spectra extracted from the corrected data product and the original data cubes, we found that without the JvM correction, the \cii\ flux of \targb\ would be overestimated by approximately 10\%. For \targa, this overestimation is less significant, likely due to the absence of pronounced extended emission in this target. The intensity maps of \cii\ emission for our targets after JvM correction are shown in Fig. \ref{fig14:comp}. We noticed that after applying this JvM correction, the extended \cii\ emission in \targb\ are more prominent with larger S/N.


\bibliography{ms}{}
\bibliographystyle{aasjournal}



\end{document}